\documentclass[aps,prfluids,nofootinbib,notitlepage,longbibliography,superscriptaddress,11pt,floatfix]{revtex4-2}
\usepackage{graphicx}
\usepackage[svgnames]{xcolor}
\usepackage{amsmath, amssymb,epstopdf,setspace}

\def\spacce#1{\hskip #1pt}
\def\drawline#1#2{\raise 2.5pt\vbox{\hrule width #1pt height #2pt}}
\def\solid{\drawline{24}{.5}\nobreak}

\def\bdash{\hbox{\drawline{5.8}{.5}\spacce{2}}}

\def\dashed{\bdash\bdash\bdash\nobreak}

\def\chndot{\hbox%
{\drawline{4.6}{.5}\spacce{2}\drawline{1}{.5}\spacce{2}\drawline{4.6}{.5}\spacce{2}\drawline{1}{.5}\spacce{2}\drawline{4.6}{.5}}\nobreak }
\def\circle{$\circ$\nobreak }
\def\linecir{\hbox%
{\drawline{8}{.5}\spacce{2}\circle\spacce{2}\drawline{8}{.5}}\nobreak}
\def\trian{\raise 1.25pt\hbox{$\scriptstyle\triangle$}\nobreak}
\def\linetri{\hbox%
{\drawline{8}{.5}\spacce{2}\trian\drawline{8}{.5}}\nobreak}
\def\dtrian{\raise 1.25pt\hbox%
{$\scriptscriptstyle\bigtriangledown$}\nobreak}
\def\rtrian{\raise 1.25pt\hbox%
{$\scriptstyle\vartriangleright$}\nobreak}

\def\squar{\raise 1.25pt\hbox{$\scriptstyle\Box$}\nobreak}

\def\diamon{\raise 1.25pt\hbox{$\scriptstyle\diamond$}\nobreak}

\newcommand{\soliddtrian}{$\blacktriangledown$\nobreak}

\def\linedtri1{\hbox{\bdash\hspace{-1.6mm}$\bigtriangleup$\hspace{-0.8mm}\bdash}\nobreak}
\def\soliddtrian1{$\blacktriangledown$\nobreak}
\def\solidrtrian2{$\blacktriangleright$\nobreak}
\def\solidltrian3{$\blacktriangleleft$\nobreak}

\def\dd{{\, \rm{d}}}
\def\dr{{\rm{d}}}

\def\bra{\langle}
\def\ket{\rangle}

\def\beq{\begin{equation}}
\def\eeq{\end{equation}}
\def\la{\label}
\def\ii{{\rm i}}

\def\r#1{(\ref{#1})}

\def\bx{\boldsymbol{x}}
\def\bu{\boldsymbol{u}}

\def\bq{\boldsymbol{q}}

\def\olz{\overline{z}}
\def\olw{\overline{w}}

\newcommand{\figpath}{./}
\begin{document}
\title{Collective organisation and screening in two-dimensional turbulence}

\author{Javier Jim\'enez}
\email[]{jjsendin@gmail.com}
\affiliation{School of Aeronautics, Universidad Polit\'ecnica de Madrid, 28040 Madrid, Spain}

\date{\today}

\begin{abstract}
Following recent evidence that the vortices in
decaying two-dimensional turbulence can be classified into small--mobile, and
large--quasi-stationary, this paper examines the evidence that the latter might be
considered a `crystal' whose formation embodies the inverse cascade of energy towards larger
scales. Several diagnostics of order are applied to the ostensibly disordered large
vortices. It is shown that their geometric arrangement is substantially more regular than
random, that they move more slowly than could be expected from simple mean-field arguments,
and that their energy is significantly lower than in a random reorganisation of the same
vortices. This is traced to screening of long-range interactions by the preferential association
of vortices of opposite sign, and it is argued that this is due to the mutual capture of
corrotating vortices, in a mechanism closer to tidal disruption than to electrostatic
screening. Finally, the possible relation of these `stochastic crystals' to fixed points of
the dynamical system representation of the turbulence flow is briefly examined.
\end{abstract}
  
\maketitle

\section{Introduction}\label{sec:intro}

This paper is part of a sequence that began as an investigation on whether an automatic
computer search of `relevant' features in decaying two-dimensional turbulence could be used
to suggest new ideas about the organisation of the flow \cite{jimploff18}. That paper was
continued in \cite{jimploff20,jotploff}, mostly from the point of view of how such
a search should be organised, and of its possible relation to automatic learning, and
eventually led in \cite{jfmploff20} into an analysis of the flow itself. It is with
the conclusions of this last paper that the present one deals.

Two-dimensional turbulence is a well-studied phenomenon, often used as an approximation to
geophysical and stratified flows in which isotropy is broken by some constraint along the
third dimension. Although truly two-dimensional turbulence is experimentally challenging, it
was one of the first turbulent flows to become accessible to direct numerical simulation,
and computational experiments were soon undertaken to test theoretical predictions
\cite{batchelor69,Joy:Mont:73,mcwilliams80,Benzi87,maltrud91}. Much of its theoretical
interest can be traced to the early observation in \cite{onsag} that a system of point
vortices in a plane could lead, under some conditions, to states of negative temperature and
to the formation of large-scale structures. We will see later that the point-vortex model is
a poor approximation to turbulence, because the former is Hamiltonian, and conserves
enstrophy (the square of vorticity) and kinetic energy, while the later is dissipative, but
the statistical mechanics of point vortices has continued to be examined in the hope that it
may be locally relevant to the dissipative case \cite{Joy:Mont:73,Mont:Joy:74,Mont:Etal:92}.
Dissipative mean-field theories based on cascades of the inviscidly conserved quantities
were developed almost simultaneously to the Hamiltonian model
\cite{kraichnan67,batchelor69}, eventually leading to the prediction of a direct cascade of
enstrophy towards smaller scales, and an inverse cascade of energy towards larger ones
\cite{kraichnan67,kraichnan71}. The latter could intuitively be related to the
negative-temperature states of \cite{onsag}.

However, simulations showed that the flow spontaneously segregates into coherent vortices
and a less-coherent background, and that the vortices interfere with the conclusions of the
mean-field representation. By the end of the 1990's there was widespread consensus that
two-dimensional turbulence is a vortex gas, following approximately Hamiltonian dynamics
punctuated by occasional mergings of vortices of like sign. The forward enstrophy cascade
proceeds by successive vortex merging and filamentation
\cite{mcwilliams90,carnevale91,Benzi92,dritsch08}, and is not described well by 
mean-field theories. The mechanism of the backwards energy cascade is less clear, although
it is not believed to be associated with the formation of larger vortices by amalgamation
\cite{Par:Tab:98,Boff:Cel:Ver:00,eyink06,xiao:09}, but the phenomenon itself has been
observed numerically, especially in forced systems kept in statistical equilibrium by a
large-scale dissipation mechanism \cite{maltrud93,borue94}. In a finite domain, and in the
absence of a dissipation mechanism, the energy accumulates at the largest available flow
scale, in a phenomenon akin to Bose-Einstein condensation \cite{smithyak93,smithyak94} and
to the predictions in \cite{onsag}. Fuller reviews of previous work on two-dimensional
turbulence can be found in \cite{tabeling02,Boff:Eck:12}.

The results in \cite{jfmploff20} extended the previous models in some interesting
directions, especially regarding the decay of unforced turbulence in the early period in
which the spectral scale of the energy has not grown enough to be affected by the simulation
domain. It was found that the largest structures of the kinetic energy take the form of elongated
high-speed jets, flanked by a subset of large vortices that move relatively slowly with
respect to the global root-mean square (r.m.s.) velocity. The cascade of energy towards
large scales corresponds to the growth of these jets. It was speculated in \cite{jfmploff20}
that the slowly moving vortex dipoles that flank the jets form a large-scale collective
structure that can approximately be described as a `crystal', and it was further speculated
that the formation of the dipoles responds to a tendency of the flow to organise into
low-energy configurations. This will be shown below to be the case, even if it may be
considered counter-intuitive in a system, like two-dimensional turbulence, which dissipates
enstrophy rather than energy, and in which the latter could be expected to `lag behind' the
decay of the former.

The questions addressed in the present paper concern this possible collective structure.
After describing in \S\ref{sec:experiments} the generation of the data and their general
properties, we examine in \S\ref{sec:energy} whether the flow really is in a low-energy
state, and whether this is connected with the presence of dipoles. Section \ref{sec:advec}
looks at the question of vortex mobility, and \S\ref{sec:lattice} and \S\ref{sec:stars}
discuss whether the organisation of the large vortices is regular enough to be considered a
crystal, even approximately.

Two related questions are addressed in the context of the previous results. The first one is
whether there is any screening between vortices of different sign, as originally suggested
by \cite{ruelle:90}, and the second one is whether the vortex `crystal' described above
could be related to the fixed-point solutions that have been invoked as organising states
for high-dimensional or turbulent dynamical systems \cite{KawEtal12}. Discussion of these
and other questions, and conclusions, are offered in \S\ref{sec:crystals} and \S\ref{sec:conclusions}.
  
\section{Numerical experiments}\label{sec:experiments}

\begin{figure}
\centerline{%
\includegraphics[width=0.80\textwidth,clip]{\figpath 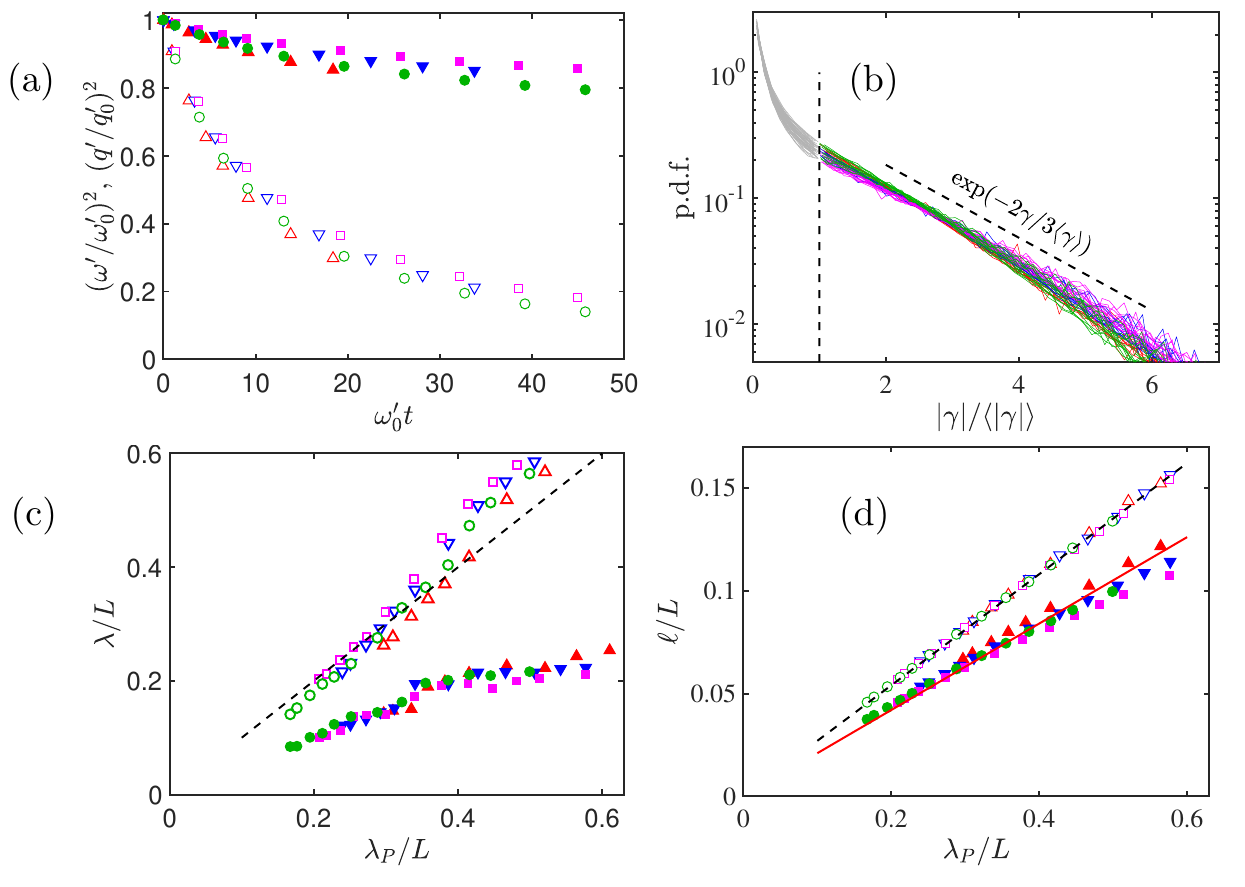}%
%
%
}%
%
\caption{%
(a) Evolution of the enstrophy (open symbols), and energy (closed symbols) of the
simulations in table \ref{tab:cases}. Symbols as in the table. Colours are: red, T512; blue,
T768; magenta, T1024; green, T1024a. They are used whenever possible throughout the paper.
(b) Probability density functions of the vortex circulation. Only vortices with
$\gamma/\bra\gamma\ket>1$ are included in the reconstructed flow fields.
(c) Comparison of the Poisson wavelength \r{eq:vortnum}, with the spectral wavelengths,
$\lambda_q$ (open symbols), and $\lambda_\omega$ (closed symbols). The dashed line is
$\lambda_q=\lambda_P$.
(d) Open symbols are the mean distance between neighbouring large vortices, and closed
symbols are the mean diameter of the large vortices, both versus $\lambda_P/L$. The dashed line
is $\bra d\ket=0.27 \lambda_P$, and the solid one is $\bra s^{1/2}\ket= 0.21\lambda_P$.
}
\label{fig:prims}
\end{figure}

The data used in this paper are mostly those in \cite{jfmploff20}, to where the reader is
referred for details. Simulations of decaying nominally isotropic two-dimensional turbulence
are performed in a doubly periodic square box of side $L$, using a standard spectral Fourier
code dealiased by the 2/3 rule. Time advance is third-order Runge-Kutta. The flow field is
defined by its velocity $\bu=(u,v)$ in the $\bx=(x,y)$ plane, and by the one-component
vorticity $\omega=\nabla\times\bu$. It is initialised with random Fourier phases and a fixed
isotropic enstrophy spectrum whose peak, approximately located at wavenumber $k_{init} =
2\pi/L_{init}$, controls the initial width of the energy-containing spectral range. The
equations are solved in vorticity--stream-function formulation, using regular second-order
viscosity, $\nu\nabla^2\omega$, and the flow is evolved for a fixed time to allow vortical
structures to establish themselves. This moment is defined as $t=0$, and denoted by a `0'
subindex in the corresponding quantities. Experiments are run as ensembles of at least 768
flows, with more samples added in some cases to improve statistics. Case T1024a, which was
not in \cite{jfmploff20}, has been included to provide a wider initial range of scales. As
the simulation proceeds, the enstrophy $\bra \omega^2\ket$, where the time-dependent average
$\bra\cdot\ket$ is taken over the full computational box and over the ensemble, decays by
approximately 70--80\%, while the kinetic energy, $\bra |\bu|^2\ket$, decreases by at most
15--20\% (figure \ref{fig:prims}a). The numerical parameters for the different cases are
summarised in table \ref{tab:cases}. The time step is controlled by a CFL condition $N
|\bu|_{max} \Delta t/3 = 0.5$, and T1024a was repeated with a twice shorter time step to
confirm the numerics.

\begin{table}
\caption{Parameters of the simulations. The size of the doubly periodic computational box is
$L\times L$. The r.m.s. vorticity $\omega_0'$, and the velocity magnitude $q_0'$, are
measured after the initial discarded transient, decaying from an initial enstrophy spectrum
whose peak is at wavelength $L_{init}$. The enstrophy and energy scales, $\lambda_{\omega
0}$ and $\lambda_{q 0}$, are defined in the text. The number of collocation points used
for dealiasing is $N\times N$. The subscript `F' refers to the end of each simulation, and
the decay time after the initial transient, $t_F$, is chosen so that
$\lambda_{qF}/L\approx 0.6$. Each case is an ensemble of $N_t$ independent
experiments. For details, see \cite{jfmploff20}\footnote{The next-to-last two columns in
this table were wrong in \cite{jfmploff20}. The present ones are believed to be correct.}.
 }
\label{tab:cases}
\begin{center}
\def~{\hphantom{0}}
\begin{ruledtabular}
    \begin{tabular}{lcccccccccccc}
      Case  & $N$ & $L_{init}/L$ & $q'_0L/\nu$  &  $N_t$ &
              $\lambda_{\omega 0}/L$ & $\lambda_{q 0}/L$ &   
              $\omega'_0t_F$ & $\omega'_F/\omega'_0$ & $q'_F/q'_0$ & Symbol \\[3pt]
\hline
      T512 & 512 & 0.05 & 4400    & 768 & 0.129 & 0.256& 23.0  & 0.50 & 0.91 &\trian \\
      T768 & 768 & 0.025 & 7800   &  768&0.108 & 0.217&   39.4 & 0.49 & 0.92 & \dtrian \\
      T1024 & 1024 & 0.033 & 11000 & 1392 & 0.095 & 0.197 & 51.3 & 0.40 & 0.92 & \squar \\
     T1024a & 1024 & 0.01 & 8900 &  2304 & 0.078 & 0.142 &  65.4 & 0.31 & 0.87 &\circle \\
   \end{tabular}
\end{ruledtabular}
  \end{center}
\end{table}
%

Natural time and velocity scales can be defined from the r.m.s. vorticity magnitude $\omega'
=\bra \omega^2\ket ^{1/2}$ and from $q' =(u'^2 + v'^2)^{1/2}$. The two length scales mainly
used in \cite{jfmploff20} are the energy spectral wavelength $\lambda_q=2\pi/k_q$, defined
by the maximum of the premultiplied energy spectrum $kE_{qq}(k)$, where $k$ is the
wavenumber magnitude, and the enstrophy wavelength, $\lambda_\omega$, similarly defined from
the enstrophy spectrum, $kE_{\omega\omega}$. Both scales increase with time, but $\lambda_q$
increases faster than $\lambda_\omega$, reflecting the energy flow towards larger scales
\cite{kraichnan67}. They are not proportional to each other, and describe different flow
features. It was shown in \cite{jfmploff20} that $\lambda_\omega$ is proportional to the
diameter of individual vortices, defined as connected objects in which the vorticity
magnitude is larger than $\omega'$, while $\lambda_q$ measures the average distance, $d$,
between closest neighbouring vortices of similar size. Our simulations are discontinued when
$\lambda_q/L\approx 0.6$, beyond which the flow is considered to enter a different decay
regime in which the energy transfer towards larger scales is substantially modified by
the finite box. As already mentioned in the introduction, vortices naturally separate into
two classes. Those whose circulation, $\gamma_i$, is weaker than the mean, $|\gamma_i|<\bra
|\gamma_i|\ket$, tend to move fast, have a power-law distribution of area and circulation,
and are only responsible for a small fraction of the total kinetic energy of the flow
\cite{jfmploff20}. Vortices stronger than the mean move more slowly, have an exponential
probability distribution (figure \ref{fig:prims}b), and are responsible for most of the
kinetic energy. This was interpreted in \cite{jfmploff20} to mean that the large vortices
form a `crystal' in approximate equilibrium, which is the subject of the present paper. From
now on, although the simulations are always run without truncation, our analysis refers to
the fraction of the flow reconstructed from the large vortices, once the smaller ones have been
deleted.

Deleting vorticity from a periodic flow  usually requires some correction, because
periodicity of the velocity field is only possible if the overall circulation vanishes,
\beq
\sum_i \gamma_i=0.
\la{eq:balance}
\eeq
When this is not satisfied, it has to be compensated by a background uniform vorticity, $-\sum
\gamma_i/L^2$. We will see below that the total kinetic energy of the flow is proportional
to the integrated squared circulation \cite{jimenez96}, so that the expected relative error
of the energy estimates in the following section is of the order of $\left( \sum\gamma_i
\right)^2/\sum \gamma_i^2$. In our reconstructed flows, it is at most $O(10^{-4})$.

\begin{figure}
\centerline{%
\includegraphics[width=0.80\textwidth,clip]{\figpath 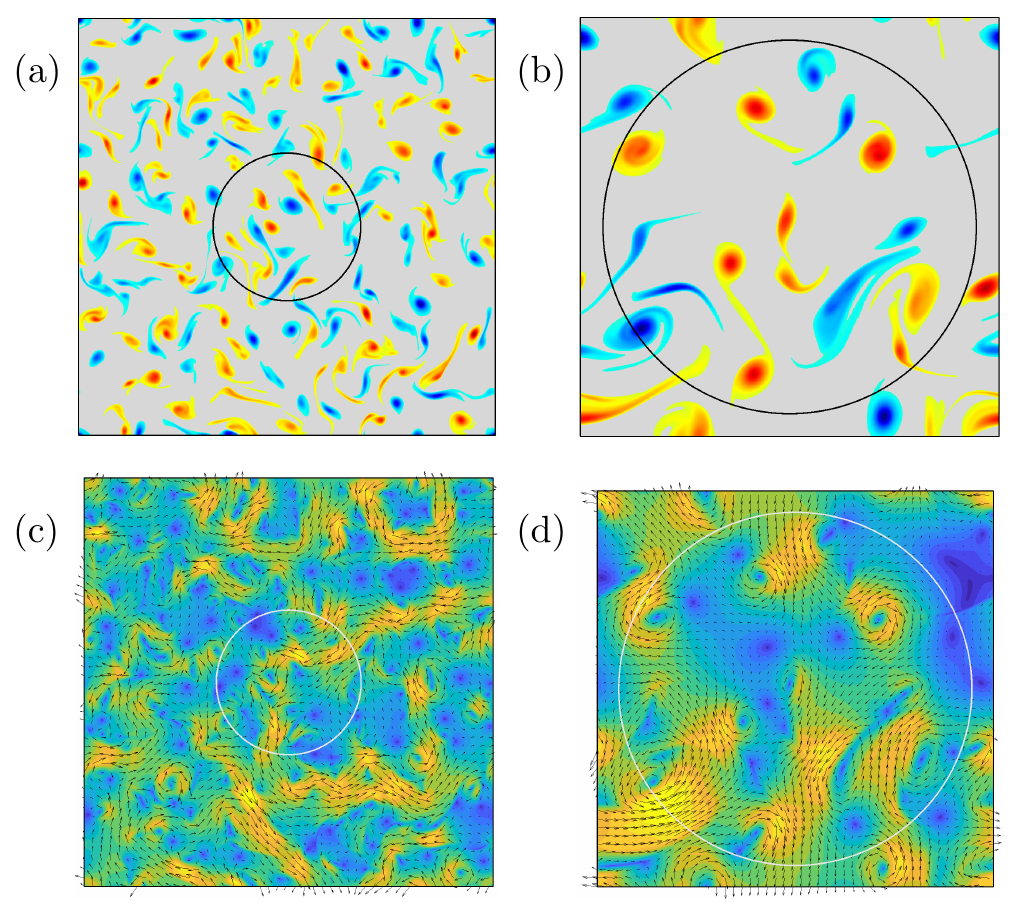}%
%
%
}%

\caption{%
(a,b) Flow vorticity synthesised by retaining only vortices whose area is larger than the
average, in one of the members of the T1024a ensemble. The circle in each case has radius
$\lambda_P$.
(a)  $\omega_0't=1.3$, $\lambda_P/L=0.18$.
(b) $\omega_0't=52.3$, $\lambda_P/L=0.45$.
(c,d) Velocity field for the flows in (a,b), respectively. The background colour is the
velocity magnitude, lighter for faster velocities.
}
\label{fig:flows}
\end{figure}

It follows from the proportionality between $d$ and $\lambda_q$ that the average number
of large vortices per flow field should decay as $(\lambda_q/L)^{-2}$. Empirically,
\beq
n_v\approx  4(L/\lambda_q)^2,
\la{eq:vortnum0}
\eeq
which decays from approximately 150 to 15 in our simulations. The last number approximately agrees with
\cite{Benzi87}, who studied similar flows at longer decay times, where $\lambda_q$ saturates to
approximately $L$, and reported that the motion of the 17 largest vortices was approximately
independent of their background. The number of small vortices is approximately twice higher than
of those larger than the mean. Because counting large vortices is a more stable operation than finding
the maximum of a noisy spectrum, especially in the late stages of the evolution in which
the definition of $\lambda_q$ depends on a few low-wavenumber harmonics, we will use in the
following a `Poisson' wavelength inspired by \r{eq:vortnum0},
\beq
\lambda_P/L = 2n_v^{-1/2}.
\la{eq:vortnum}
\eeq
The relation between $\lambda_P$ and the two spectral wavelengths is shown in figure
\ref{fig:prims}(c). Figure \ref{fig:prims}(d) shows its relation to the mean distance
between large vortices, and to their diameter. It is not surprising that $\lambda_P$ is
proportional to the distance between vortices, because this is how it is defined. In fact,
assuming that the vortices are a Poisson-distributed set of points with density $n_v/L^2$, 
their average separation would be $\bra d\ket =0.5L/n_v^{1/2}=0.25\lambda_P$
\cite{jfmploff20}. This linear dependence, and even its coefficient, are well satisfied in
figure  \ref{fig:prims}(d).

More interesting is that $\lambda_P$ is also linearly related to the vortex diameter, $\bra
s^{1/2}\ket$, where $s$ is the area of the thresholded vortices. This was not the case for the full
vortex system analysed in \cite{jfmploff20}, where the diameter was found to be proportional
to $\lambda_\omega$, which is not proportional to $\lambda_q$. Figure \ref{fig:prims}(d)
implies that, even if the decay of the full flow is not self-similar in this way, the decay
of the system of large vortices is truly self-similar, with a vortex diameter proportional
to the separation. Two snapshots of the temporal evolution of the large-vortex component of
a turbulent flow are shown in figure \ref{fig:flows}. The circles in the figure have radius
$\lambda_P$, and \r{eq:vortnum} implies the number of vortices within them should stay
approximately equal to $4\pi$. The figure also displays the velocity field of these two
flows, clearly showing the high-velocity streams mentioned in the introduction
\cite{jfmploff20}.

Although not discussed in detail in this paper, reference will occasionally be made to
simulations of Hamiltonian systems of point vortices. Their characteristics and numerical
details are described in Appendix \ref{sec:pointv}.

\section{The kinetic energy}\la{sec:energy}

\begin{figure}
\centerline{%
\includegraphics[width=0.80\textwidth,clip]{\figpath 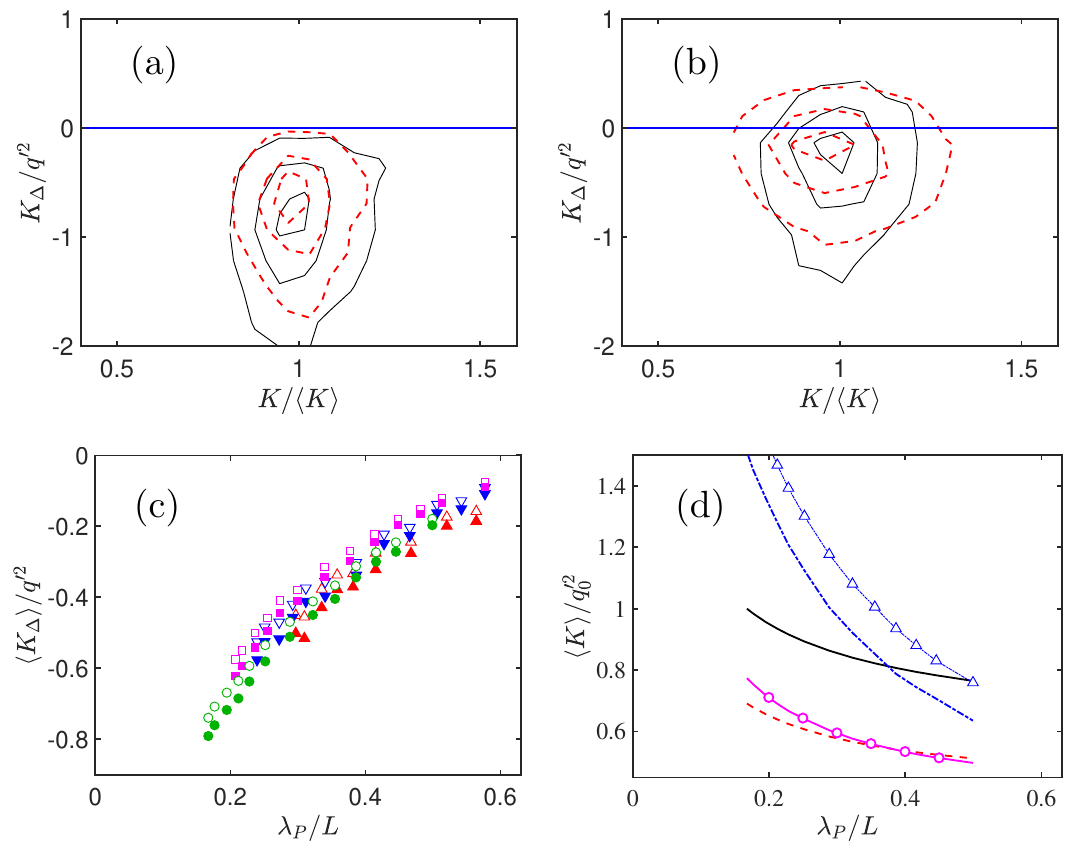}%
%
%
}%
\caption{%
(a) Joint probability density function of the kinetic energy difference, $K_\Delta=K-K_s$,
between the extended-vortices flows, $K$, and the same flow in which the vortex positions
are randomised, $K_s$. T1024a at the time of figure \ref{fig:flows}(a).
(b) As in (a),  at the time of figure \ref{fig:flows}(b).
(c) Ensemble average of the randomising energy, versus $\lambda_P$.
Symbols as in table \ref{tab:cases}. 
In (a-c), open symbols and solid lines are the turbulence vortices, and closed symbols and
dashed lines are the point-vortex approximation. Contours in (a,b) contain 10\%, 50\% and
90\% of the probability.
(d) Kinetic energy evolution. \solid, Total kinetic energy; \dashed, energy from all
vortices; \linecir, energy from large vortices; \chndot, energy from scrambled large vortices;
\linetri, self-energy approximation to the scrambled large vortices.
T1204a.
}
\label{fig:energy}
\end{figure}

The clearest indication that the arrangement of the large vortices is not random is figure
\ref{fig:energy}, which compares the total kinetic energy of an ensemble of large-vortex reconstructions of
turbulent flows with that of another ensemble in which the position of the same vortices is randomised. Figures
\ref{fig:energy}(a,b) are joint probability density functions (p.d.f.) of the kinetic energy of the members of the ensemble,
\beq
K=L^{-2} \int_{L^2} |\bu|^2 \dd^2\bx,
\la{eq:Kdef}
\eeq
versus the energy required for randomisation, $K_\Delta=K-K_s$. Particularly at the earlier
evolution time in figure \ref{fig:energy}(a) (one of whose members is represented in figure
\ref{fig:flows}a), the scrambling of the vortex position almost always leads to higher
energies, although the effect weakens as the flow evolves towards fewer vortices in figures
\ref{fig:energy}(b) and \ref{fig:flows}(b). Some caution is required in interpreting this
result, because the vortices occasionally overlap after being rearranged, modifying the
total enstrophy and the energy. The overlap is of the order of 7-10\% of the vortex area,
and, while the effect on the enstrophy is statistically neutral (not shown), that on the
kinetic energy is not. Its magnitude is tested in figure \ref{fig:energy}(a,b) by repeating
the scrambling experiment after substituting each vortex by a compact `point' core with the
same circulation, located at its centre of gravity (see appendix \ref{sec:pointv}). The
overlap then decreases by a factor of 2 to 10 (see appendix \ref{sec:overlap}), depending on the
case, but the effect of the scrambling is still to increase the energy. Figure
\ref{fig:energy}(c) summarises the energy increase for all the experiments and
evolution times. The best collapse is obtained by plotting it against $\lambda_P/L$, or,
equivalently, against the number of vortices in the periodic domain.

The kinetic energy of a sparse zero-circulation system of two-dimensional small vortices is
\cite[\S 7.3]{bat67},
\beq
2\pi K  \approx 
\left(\sum_i \gamma_i^2\right)\log (\delta/\ell) + \sum_i \sum_{j\ne i} \gamma_i \gamma_j \log( r_{ij}/\ell)  +\ldots,   
\la{eq:vener}
\eeq
where $r_{ij}$ is the distance between vortices, $\ell$ is an arbitrary length scale, and
$\delta$ is of the order of the radius of the vortex cores. The first term in the right-hand
side of this equation is the self energy, and is independent of the vortex position. It
derives from the $1/r$ velocity in the neighbourhood of the cores, and diverges as
$\log(\delta)$. The second term in \r{eq:vener} is the interaction energy. It is independent
of $\delta$, and only depends on the relative vortex position.

When \r{eq:vener} is modified to take into account the spatial periodicity of the
computational box, the argument of the logarithms gains a Jacobian elliptic function
\cite{Oneil:89}, which does not change the short-range interactions but introduces $\ell=L$
as a natural length scale. With this choice, the self-energy of the cores increases with
decreasing core size, and is 2--3 times larger for the point-vortex model than for the
extended vortices of the original turbulence. This is the reason why the abscissae in figure
\ref{fig:energy} have to be scaled with the individual average for each system, but the
figure shows that the interaction energy spent in randomisation is the same for both flows,
which is why their ordinates can be scaled with the same energy ($q'^2$ of the turbulent
flow).

One of the most intriguing features of figure \ref{fig:energy}(c) is that the magnitude of
the available interaction energy is initially of the same order as the total kinetic energy
of the flow, but that it almost vanishes at the end of the simulations. This appears to contradict
the result in figure \ref{fig:prims}(a) that the kinetic energy stays approximately constant
during the decay, and the implication is that most of the interaction energy is never
expressed in the flow. The energy redistribution mechanism responsible for keeping vortices
in a reduced-energy state is conservative, unrelated to dissipation, and the position of the
vortices in a turbulent flow is always far from random.

Consider figure \ref{fig:energy}(d), which shows the evolution of several components of the
energy of the flow as it decays. The black solid line is the total kinetic energy, which
decays by about 20\%. The same is approximately true of the part of the kinetic energy
contained in the large vortices, which closely tracks the total. The approximately 20--25\%
separating the two energies is presumably contained in the incoherent background. Including
the vortices that we have classified as small (red dashed line) does not change the
picture \cite{jfmploff20}. They actually subtract some energy from the flow during most of
the evolution. The largest energy reservoir is in the set of randomised vortices (blue
chain-dotted line), which is almost exclusively self energy. The triangles in figure
\ref{fig:energy}(d) are an approximation to the kinetic energy that essentially treats the
first term in \r{eq:vener} as a sum of independent variables (see appendix
\ref{sec:lattice_ener}). It represent the data well, especially taking into account that the
turbulence vortices are far from circular, and that the definition of
$\delta=\bra(s/\pi)^{1/2}\ket$, is ambiguous.

\begin{figure}
\centerline{%
\includegraphics[width=0.80\textwidth,clip]{\figpath 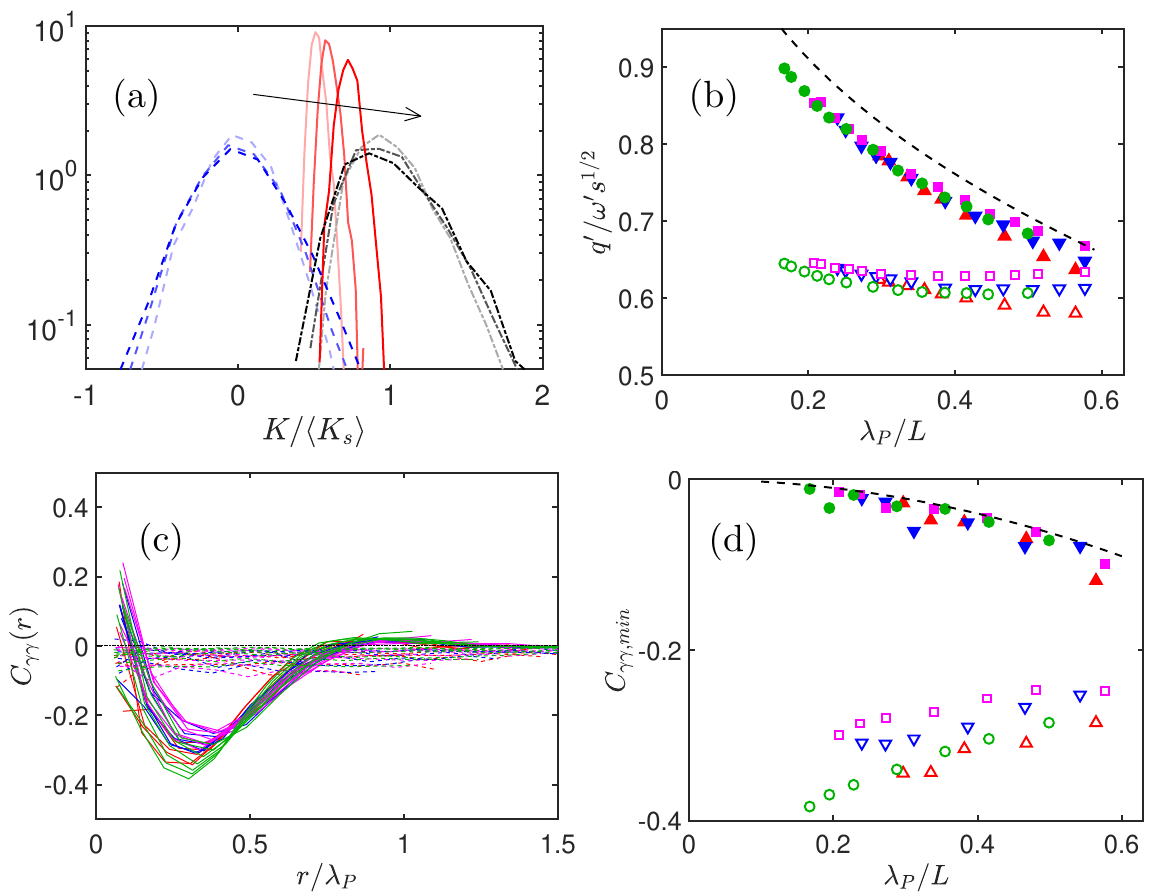}%
%
}%
\caption{%
(a) One-dimensional p.d.f. of the kinetic energy of the reconstructed flows. \solid,
Unscrambled vortices; \chndot, scrambled; \dashed, difference between two scramblings of the
same flow. T1024a. Time increases from light to dark, $\omega'_0 t=$ 1.3, 19.6, 52.3, and all
energies are normalised with the mean scrambled energy at each time.
(b) Normalised velocity intensity. Symbols as in table \ref{tab:cases}. Open symbols are
unscrambled flows; solid ones are scrambled. The dashed line is the logarithmic correction
\r{eq:gaussn}.
(c) Correlation of the vortex circulation as a function of their distance. All cases and
times. Colours as in (b). \solid, unscrambled; \dashed, scrambled.
(d) Minimum of the correlations in (c). Symbols as in (b), but the dashed line is \r{eq:cggminus}.  
}
\label{fig:energy1D}
\end{figure}

Figure \ref{fig:energy1D}(a) examines the energy distribution in more detail. The solid red
lines are p.d.f.s of the kinetic energy for the ensemble of turbulent flows, plotted at
three evolution times that increase in the direction of the arrow and of the darker shades
of colour. In each case, the energy is normalised with the average energy of the ensemble of
scrambled flows at the same time, and the rightward drift of the p.d.f. corresponds to the
upward trend of the open symbols in figure \ref{fig:energy}(c). The black dashed lines in
figure \ref{fig:energy1D}(a) are p.d.f.s of the energy of the scrambled flows, scaled with
their own mean. They change little with time, and we have seen that they mostly include the
self-energy of the vortices. It could be expected that, since the vortex positions are
randomised, and since the velocity at each point of the flow is the sum of the contributions
of the different vortices, the central-limit theorem would result in a Gaussian p.d.f., but
\cite{jimenez96} examined this limit, and concluded that the $1/r$ velocity distribution due
to each core in a system of random compact vortices is marginal for the application of the
central-limit theorem. Although the p.d.f. of the resulting velocity would tend to Gaussian
for a very large number of vortices, the rate of convergence is too slow to be useful in any
practical sense. As a consequence, the distributions in figure \ref{fig:energy1D}(a) are
skewed, as in \cite{jimenez96}. Finally, the blue chain-dotted curves in figure
\ref{fig:energy1D}(a) are the interaction energy, which has been isolated by subtracting the
energy of two different randomisations of the same flow, $K_{\Delta\Delta}
=(K_{s1}-K_{s2})/\sqrt{2}$. These distributions are symmetric with zero mean, by
construction, and, at least for the flows represented here, collapse well when scaled with
the mean scrambled energy. Their standard deviation is of the order of 25--30\% of this
mean, proving that there is enough energy in the vortex interactions to explain the energy
required for scrambling.

It is significant that the distribution of the energy within the ensemble of real flows is
much narrower than in the scrambled ones, implying again that turbulence only explores a
small subset of the possible vortex distributions. Figure \ref{fig:energy1D}(b) shows the
scaling of the kinetic energy with the enstrophy. Dimensionally, $q'\sim \omega' \bra s\ket
^{1/2}$, where $\bra s\ket$ is the average vortex area, but there is a logarithmic correction
(appendix \ref{sec:lattice_ener}), with the same origin as in figure \ref{fig:energy}(d),
\beq
q'^2/\omega'^2 \bra s\ket \sim \log(\ell_0/R),
\la{eq:gaussn}
\eeq
where $\ell_0$ and $R$ are outer and inner limits, respectively, of the region over which
the velocity induced by a vortex core decays as  $1/r$. The inner scale is
always of the order of the vortex radius $R\sim \bra s^{1/2}\ket $, but $\ell_0$ depends on the vortex
distribution. For a random distribution of vortices in a box of side $L$ we can assume
$\ell_0\approx L/2$ and, since we saw in figure \ref{fig:prims}(d) that $\bra s^{1/2}\ket \approx
\lambda_P/4$, the right-hand side of \r{eq:gaussn} is approximately $\log(2L/\lambda_P)$.
The ratio $q'/\omega' \bra s ^{1/2}\ket$ is plotted in figure \ref{fig:energy1D}(b), where
the difference between the scrambled and unscrambled flows is equivalent to the one in
figure \ref{fig:energy}(c). It is plotted against $\lambda_P/L$, and the logarithmic
correction \r{eq:gaussn} is included as a dashed line. It describes the trend of the
randomised flows well. Interestingly, the normalised energy of the original turbulent flows
satisfies the uncorrected constant scaling much more closely than the scrambled ones,
consistent with $\ell_0/R\approx 1.5-2$. This suggests that the turbulence vortices suppress  each
other's $1/r$ velocity field beyond distances of the order of $d/\lambda_P\approx 0.3-0.4$,
which is also the typical distance between nearest neighbours.

That neighbouring vortices are not randomly distributed is confirmed by figure
\ref{fig:energy1D}(c), which displays the correlation coefficient of vortex circulation as a
function of distance,
\beq
C_{\gamma\gamma}(r)=\bra \gamma_i \gamma_j\ket/\bra\gamma_i^2\ket,
\la{eq:cgg}
\eeq
where $r=\bra r_{ij}\ket$, and the average is taken over the ensemble and over all the
$(i\ne j)$ combinations. Since, on average, there is the same number of vortices of each
sign, one could expect this correlation to vanish. This is approximately true for the
scrambled flows, plotted in the figure as dashed lines, but turbulent flows behave
differently, and the correlation reaches a negative minimum at $r\approx 0.33\lambda_P$.
This is close to the distance between neighbouring cores (figure \ref{fig:prims}d), and
implies that near neighbours tend to be of opposite sign. We noted when discussing
\r{eq:vortnum} that the distribution of distances among neighbouring vortices is consistent
with a Poisson distribution, and the same is found in \cite{jfmploff20}, but, when the
vortices in that paper are grouped into closest pairs, the number of counter-rotating
dipoles is found to be approximately 50\% larger than that of co-rotating pairs. This
effect disappears when the vortex position is randomised.

The minimum values of the correlations in figure \ref{fig:energy1D}(c) are collected in
figure \ref{fig:energy1D}(d). Those from the turbulent flows increase slowly with increasing
$\lambda_P/L$, but those from the scrambled ones do the opposite, and it is intriguing that
they are always negative, even if they are independent of the distance $r$. This would seem
to imply that some structure remains in the randomised flows, but the explanation is
probably simpler, and, as we shall see below, significant. Consider a flow with $n_v$
identical vortices of balanced sign. Each vortex sees $n_v/2$ vortices of opposite sign, for
which $\gamma_i\gamma_j<0$, but only $n_v/2-1$ vortices for which $\gamma_i\gamma_j>0$,
discounting itself. Using \r{eq:vortnum}, the resulting average is
\beq
C_{\gamma\gamma}\approx -1/n_v = -0.25 (\lambda_P/L)^2.
\la{eq:cggminus}
\eeq
This line has been added to figure \ref{fig:energy1D}(d), and represents the data  well.
 
Screening of the far-field electrostatic potential by the redistribution of charges is 
well known in plasmas and electrolytes, where a charged particle induces a shell
of charges of opposite sign that shelters its effect at long distance \cite[\S
75]{lan:lip:5}. This sheltering reduces the total potential energy, and it was speculated in
\cite{ruelle:90} that a similar effect could be present in turbulence. There is little
empirical or experimental evidence either way, although \cite{ishetal06,Davidson:11}
observed that the absence of long-range pressure correlations in simulations of
three-dimensional isotropic turbulence could be due to screening, and generalised the
argument to other three-dimensional flows. On the contrary, \cite{jimenez96} found that a random
vortex distribution describes well the second-order moments of decaying two-dimensional
turbulence, and interpreted this as evidence for the absence of screening. The reason for
this discrepancy with the present results is not clear, but the two simulations are
difficult to compare, because \cite{jimenez96} uses fourth-order hyperviscosity as a
dissipation model. Hyperviscous vortices are typically smaller than viscous ones, and the
area fraction covered by the vortices in that simulation is an order of magnitude smaller
than in the present one. It is likely that $q'$ in \cite{jimenez96} is mostly self-energy

On the other hand, electrostatic screening is quite different from vortex rearrangement, as
already noted in \cite{ruelle:90}. The electrostatic Debye--H\"uckel screening length arises 
from a balance between the potential energy of the charge distribution and the kinetic
energy of thermal agitation \cite{lan:lip:5}. The rationale for vorticity screening would
also be to minimise total energy, and it was already speculated in \cite{jfmploff20} that
the predominance of dipoles in simulations could be an energy effect. But, although it is
clear from \r{eq:vener} that tight pairs of cores of opposite sign decrease the interaction
energy with respect to those of the same sign, this is due to their effect on the kinetic
energy in the far field. There is no potential energy among vortices, and calculations of
the evolution of point vortex systems (see appendix \ref{sec:pointv}) reveal
no screening effect. Their correlation coefficient \r{eq:cgg} remains equal to zero up to
an inner limit determined by the temporal resolution of the numerical scheme, and the
distribution of distances among close vortices is consistent with Poisson statistics, both
for pairs of the same sign and for dipoles.

In fact, the evolution of inviscid point vortices and the decay of two dimensional
turbulence are fairly different problems, because, as mentioned in the introduction, the
former is a Hamiltonian system that conserves energy, momentum and, by default, enstrophy,
while the latter is dissipative. But there is persuasive evidence that they are not
completely unrelated, and \cite{Benzi87} showed that substituting points for the largest
vortices in decaying turbulence reproduces well their motion for short times, while
\cite{Benzi92} showed that adding a simple merging rule for like-signed vortices, designed
to conserve energy while dissipating enstrophy, extends the validity of the model to longer
decays. On the other hand, if screening were just a local energy effect, it should be
visible in the Hamiltonian simulations, which we have seen not to be the case. A fair amount
of work has gone into the equilibrium solutions of sets of point vortices, from vortex
crystals that are steady in some frame of reference \cite{aref02} (see also
\S\ref{sec:crystals} below), to maximum-entropy distributions which are only statistically
stationary \cite{Joy:Mont:73,Mont:Joy:74}. These last two papers centre on the statistical
mechanics of sets of point vortices, in terms of an interaction energy measured with respect
to the random distribution. They find that, when the interaction energy is positive, the two
vortex signs separate into large coherent regions of opposite circulation, as in the
negative `temperature' states in \cite{onsag}. For negative energies, equivalent to positive
temperatures, the two signs stay mixed. It follows from figures \ref{fig:energy} and
\ref{fig:energy1D} that the latter is the regime of our decaying flows, except perhaps for
distances of the order of the separation between immediate neighbours.

The closest equivalent to screening in the context of vortex flows
is shear sheltering \cite{hunt:carr:90,Terry:00,hunt:eam:wes:06}, in which a vortex
approaching a strong vorticity interface deforms it in such a way as to cancel the
long-range effect of the vortex. This has been invoked, for example, to explain the
persistence of strong vortices in two-dimensional turbulence \cite{Terry:00}, and of regions
of very different turbulence intensities in three-dimensional turbulence
\cite{hunt:eam:wes:06}, but the effect is restricted to the interaction of vorticity
structures of very different magnitude. Thus, while sheltering might be involved in the
separation of vortex cores into large and small, it is unlikely to be the reason
why the strong vortices rearrange themselves into low-energy configurations.

More promising is an extension of the argument leading to \r{eq:cggminus}, and it may be
relevant that, if this equation is extrapolated to $\lambda_P \approx L\, (n_v\approx
5)$, it approaches $C_{\gamma\gamma,min}\approx -0.25$, which is of the same order as the
observed minima for the turbulence data. This, together with the observation that the
negative correlations are restricted to distances of the order of the nearest vortex
neighbours, suggests that the vortex circulation tends to be in balance over neighbourhoods
of $O(\lambda_P)$, in such a way that each vortex sees an excess of neighbours of the
opposite sign. There are several ways in which this might be implemented. The simplest one
is probably vortex amalgamation. When two point vortices are drawn together by a
multi-vortex interaction, they tend to stay together for some time, among other
reasons because separating them requires exchanging an energy  that may not be locally
available. This is true for point vortices of any sign, but the long term evolution of
pairs of extended patches depends on their relative sign. Corrotating pairs tend to merge
\cite{meunier05}, so that one of the partners disappears from the local count, while
counter-rotating dipoles are stable \cite{Flierl:80,mcwilliams80}. This would explain the
prevalence of dipoles over co-rotating pairs, and the local correlation minimum in turbulence
simulations, as well as the absence of both effects in point-vortex systems.

Vortex amalgamation depends on the ratio of the vorticity to the strain that the 
partners induce on each other \cite{moo:saff:71}, which, for approximately similar vortices,
only becomes relevant at distances of the order of the vortex radius
\cite{moo:saff:75}. This would be consistent with the observation in
\cite{jfmploff20} that the distribution of vortices in decaying turbulence satisfies Poisson
statistics except at distances of the order of $\lambda_\omega$, proportional to the vortex
diameter. On the other hand, figure \ref{fig:energy1D}(c) suggests that the circulation is
correlated over distances of the order of $\lambda_P$, which is of the order of the distance among
vortices, but we saw in figure \ref{fig:prims}(d) that distance and diameter are
proportional to each other for the large vortices. In fact, mutual shredding provides a
plausible mechanism to enforce this proportionality \cite{moo:saff:75}. Note that, in this
interpretation, screening in turbulence would be closer to the Roche mechanism by which
celestial bodies are destroyed by tidal forces in a strong gravitational field
\cite{chandra:63}, than to the charge redistribution in plasmas.

\section{Vortex mobility}\la{sec:advec}

\begin{figure}
\centerline{%
\includegraphics[width=0.8\textwidth,clip]{\figpath 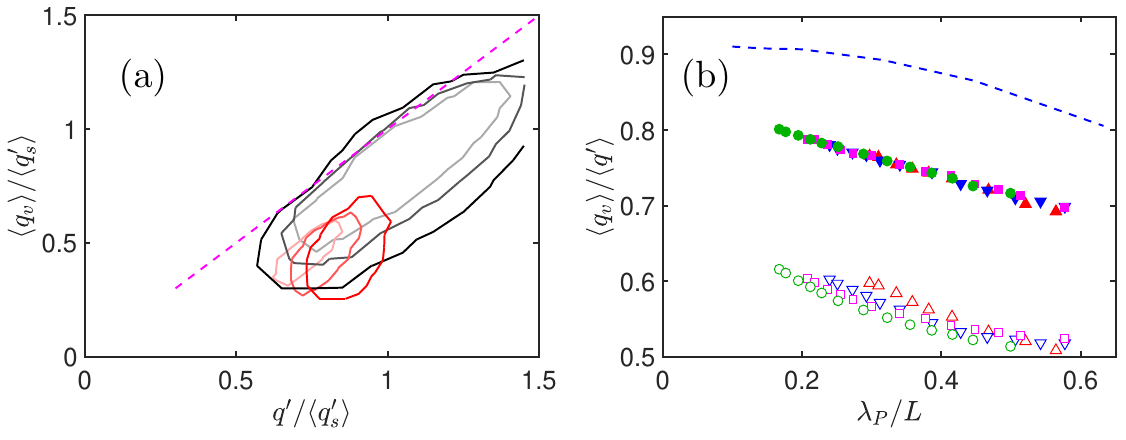}%
%
}%
\caption{%
(a) Joint p.d.f. of the r.m.s. velocity and mean vortex mobility. Red lines are for
large-vortex turbulence; black are for scrambled positions. Time increases from light to
dark, $\omega'_0 t=$ 1.3, 19.6, 52.3, and all the energies are normalised with the mean
scrambled energy at each time. Contours contain 99\% of the probability mass.
(b) Ratio between mobility and r.m.s. velocity. All cases. Symbols as in table
\ref{tab:cases}. Open symbols are unscrambled flows; solid ones are scrambled, and the
dashed line is for the Hamiltonian evolution of sets of point vortices with an exponential
distribution of circulation. The $\lambda_P$ for the point sets is defined using
\r{eq:vortnum}.
}
\label{fig:advec}
\end{figure}

The second interesting property of the system of large vortices is that they move more
slowly that would be expected from the root-mean square velocity of the flow
\cite{jfmploff20}. In particular, if we define the mobility of a vortex core as   
\beq
q_v = s^{-1} \left| \int_s \bu \dd^2 \bx \right| 
\la{eq:mobility}
\eeq
where the integral is taken over the vortex support, it would appear that $\bra q_v\ket$
should be approximately equal to $q'$. Figure \ref{fig:advec}(a) shows that this is not the
case. The black contours in the figure are joint probabilities of r.m.s. velocity versus
mobility in randomised vortex systems. As could be expected, flows with lower kinetic energy
also induce a lower mobility in their cores, but real turbulence, represented by the red
contours, not only has a narrower distribution of energy than the scrambled flows, as in
figure \ref{fig:energy1D}(a), but also lies at the lower edge of the range of mobilities.
Considering that the vortices that have been used for the black contours are the
same ones used for the red ones, and that the contours enclose 99\% of the probability
mass in both cases, the vortex distributions in turbulence correspond to very improbable
geometric arrangements. For example, the distance between two probability distributions
$p_1(\bq)$ and $p_2(\bq)$, in a parameter space $\bq$, can be quantified by the Hellinger
norm \cite{nikulin01}, defined by
\beq
H^2(p_1, p_2) = \tfrac{1}{2} \int (\sqrt{p_1}-\sqrt{p_2})^2 \dd \bq,
\la{eq:Hell}
\eeq
which vanishes for $p_1=p_2$, and reaches its maximum, $H=1$, for disjoint distributions. In
the case of the scrambled and unscrambled distributions in figure \ref{fig:advec}(a), it
varies from $H\approx 0.97$ at early times, when there are many vortices, to $H\approx 0.75$ at the
end of the simulations.
 
Even if the scrambled flows in figure \ref{fig:advec}(a) tend to have $q_v<q'$, probably
because \r{eq:mobility} acts as a filter that smooths the velocity fluctuations, figure
\ref{fig:advec}(b) shows that the low mobility of the turbulence vortices goes beyond this
effect. The closed symbols in this figure are the ratio between the mobility and the
fluctuation intensity in scrambled flows. It is of the order of 0.75. The open symbols
are for turbulence, and, even if the vortices in both cases are the same, and therefore
subject to similar filtering effects, the turbulent flows have smaller ratios, of the order
of 0.55. Both ratios decrease for decreasing number of vortices, in what is probably another
manifestation of the self-exclusion effect in \r{eq:cggminus}. The dashed line in figure
\ref{fig:advec}(b) is computed for Hamiltonian systems of point vortices, which, as we saw
above, have very different dynamics from turbulence. In addition, their mobility is a true
advection velocity, rather than an average such as \r{eq:mobility}. Even so, and even if
their mobility ratio is closer to unity than in the case of extended vortices, it also
decreases with the number of vortices. The simplest explanation, as in \r{eq:cggminus}, is
that the self-induced velocity of each vortex is included in $q'$ but not in $q_v$, and that this
difference becomes more important as the number of vortices decreases.

A possibility that needs to be discounted is that the low mobility of the turbulence
vortices may be due to geometric jamming, since we saw in figure \ref{fig:prims}(d) that the
inter-vortex distance is of the same order as their diameter. This is not tested by the
randomisation of the vortex position in figure \ref{fig:advec}, because we saw above that
this scrambling leads to some vortex overlap, which makes jamming partly irrelevant, but it
can be tested by randomly flipping the sign of the vortices, without modifying their
vorticity distribution or their shape. This operation respects the vortex geometry and would
be as jammed as the original flow field, but the results (not shown) are indistinguishable
from those for scrambled positions in figure \ref{fig:advec}(b).

\section{The vortex lattice}\la{sec:lattice}
\begin{figure}
\centerline{%
\includegraphics[width=0.80\textwidth,clip]{\figpath 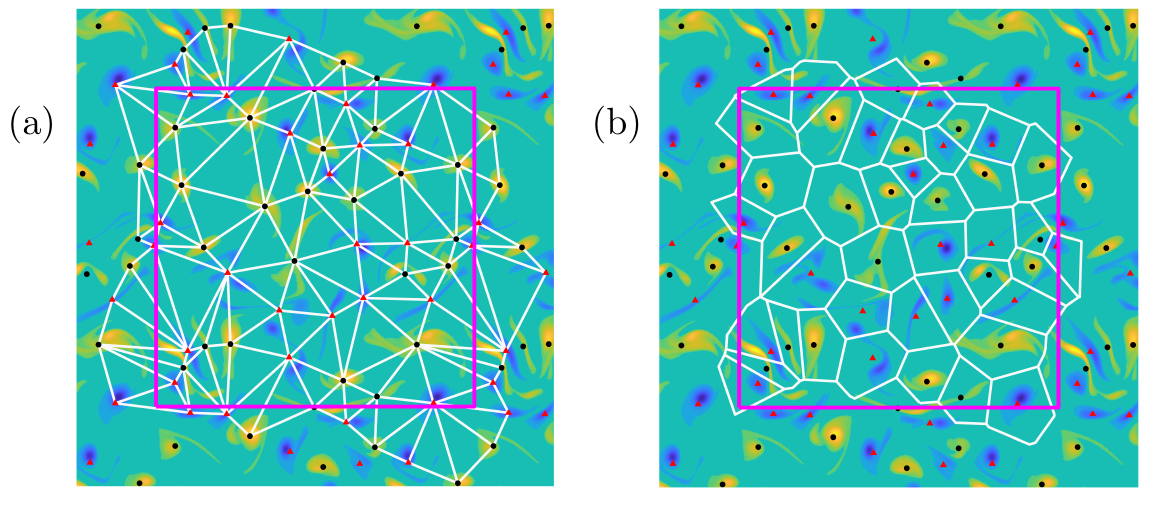}%
%
}%
\caption{%
(a) Delaunay triangulation of the lattice of vortex centres. Black dots are positive
vortices; red dots are negative. The purple frame is the periodic computational box, and
only triangles containing a vertex within the box are plotted. T512.
$\omega'_0t=2.8, \,\lambda_P/L=0.30$.
(b) The dual Voronoi tessellation of the same field.
}%
\label{fig:delaunay}
\end{figure}

\begin{figure}
\centerline{%
\includegraphics[width=0.80\textwidth,clip]{\figpath 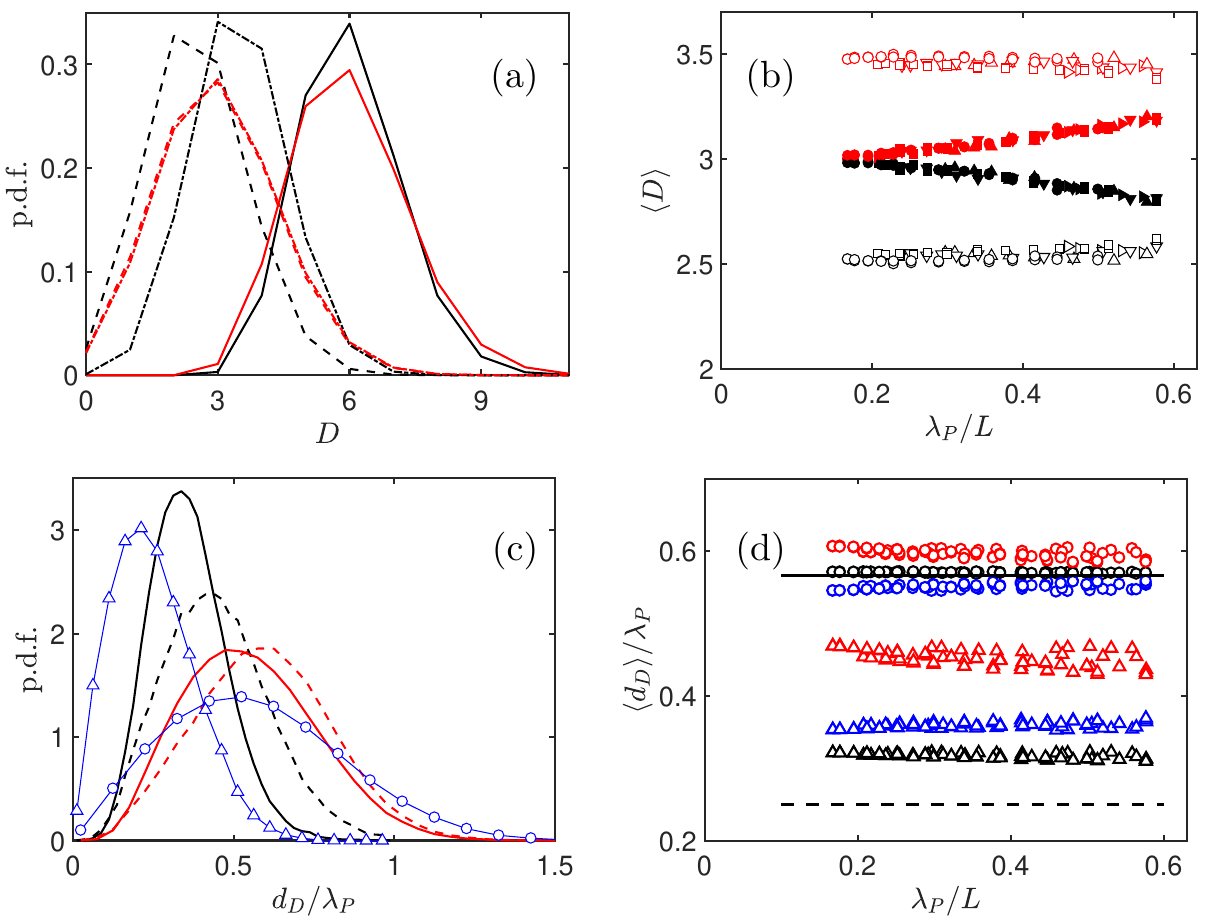}%
%
}%
%
\caption{%
Properties of Delaunay lattice neighbours.
(a) P.d.f. of the number of neighbours. \solid, All neighbours; \dashed, same circulation
sign; \chndot, opposite sign. Red lines are for scrambled vortex positions. T1024a, as in
figure \ref{fig:flows}(a).
(b) Mean number of neighbours as function of number of vortices per field. Black symbols are
corrotating with the reference vortex. Red ones are counterrotating. All cases. Symbols as in table
\ref{tab:cases}. Open symbols are unscrambled flows; solid ones are scrambled circulations.
(c) P.d.f. of the distance between Delaunay neighbours. T2014a, as in figure
\ref{fig:flows}(b). Red lines refer to all neighbours; black lines, to the closest one in
each neighbourhood; \solid, counterrotating with the reference vortex; \dashed, corrotating.
Blue lines with symbols used scrambled vortex coordinates. \circle, all neighbours; \trian,
closest one.
(d) Average distance from a vortex to its Delaunay neighbours. \circle, all neighbours;
\trian, closest one. Red symbols only include corrotating vortices; blue are
counterrotating, and black include both signs. The two horizontal lines are for scrambled
coordinates. \solid, all neighbours; \dashed, closest one.
}
\label{fig:latt}
\end{figure}

We next address what makes the distribution of vortices in turbulent
flows atypical. We will focus on the organisation of the centres of gravity of the vortices
within their local neighbourhood, since the two previous sections suggest that the vortex
distribution over long distances is essentially random.
 
The conjecture at the end of the \S \ref{sec:energy} about the structure of the vortex
neighbourhood can be tested by analysing the lattice formed by the vortex centres. Consider
the example in figure \ref{fig:delaunay}, where the Delaunay triangulation and its
associated Voronoi tessellation have been superimposed on the centres of gravity of the
vortices within the periodic domain marked by the purple frame \cite{Stoyan87}. The Delaunay
triangulation is the `most regular' triangulation of neighbouring lattice points, and
defines lattice neighbourhoods. The Voronoi tessellation is formed by the sets of flow
points which are closest to each vortex centre, and defines flow neighbourhoods.

The lattice neighbourhood of a vortex centre can be defined as the set of centres connected
to it by a Delaunay edge, and the conjecture in  \S \ref{sec:energy} can be expressed as
that the lattice neighbourhood of a vortex contains, on average, less vortices corrotating
with it than counterrotating ones. This would automatically be true for any reasonably
homogeneous lattice by the argument in \r{eq:cggminus}, and a second question is whether the
observed effect is stronger than could be expected from a random point set. This is tested
in figure \ref{fig:latt}(a), which shows the probability distribution of the degree, $D$
(number of neighbours connected to a given vortex), as well as of the number, $D_+$ and
$D_-$, of corrotating and counterrotating neighbours, respectively. It is known that $\bra
D\ket=6$ for any planar triangulation \cite{Stoyan87}. This is a topological property,
independent of flow structure, but it is clear from figure \ref{fig:latt}(a) that $\bra
D_-\ket>\bra D_+\ket$. The average values of these partial degrees are given in figure
\ref{fig:latt}(b). The open symbols are $D_+$ and $D_-$; they depend little on the
particular case or on the evolution time, and reflect the average loss of one corrotating
vortex from each neighbourhood. To test whether this loss is specific to turbulence, the
triangulation is repeated after randomising the circulation of the vortices. The results are
included as solid symbols in figure \ref{fig:latt}(b). They show no difference between the
two vortex signs at short evolution times, when the flow has many vortices, and drift
towards the unscrambled results when the number of vortices decreases at later times. The
result of randomising the vortex position instead of the circulation are essentially the
same. The origin and magnitude of this drift is the same as in figure \ref{fig:energy1D}(d)
and in \r{eq:cggminus}.

Figures \ref{fig:latt}(c,d) display the p.d.f. and mean value of the distance from a vortex
to its Delaunay neighbours, $d_D$, separated into corrotating and counterrotating ones. The red
lines in figure \ref{fig:latt}(c) and the circles in \ref{fig:latt}(d) refer to all the
neighbours, while the black lines and the triangles refer to the closest one. In both cases
they are compared with the result from fields in which the vortex positions are
scrambled. The results depend relatively little on the evolution time or on the number of
vortices. In all cases, the corrotating neighbours are farther from the reference vortex
than the counterrotating ones, in agreement with the `tidal capture' model discussed in
\S\ref{sec:energy}. This effect is weak in the case of all neighbours, $\bra
d_D\ket/ \lambda\approx 0.59$ and 0.55, respectively, compared to 0.57 in the randomised
case, but much clearer in the case of the closest neighbour, $\bra d_D\ket/ \lambda\approx
0.44$, 0.36 and 0.25, respectively. Also striking is that the turbulence p.d.f.s are
narrower than the randomised ones, reflecting a more `rigid' geometry that could be
interpreted as supporting the description of the lattice as a `crystal'. The main
difference between turbulence and the randomised field is at short distances, where the
vortices do not approach each other beyond $d\approx 0.25 \lambda_P$, of the order of their
diameter (figure \ref{fig:prims}d). This is again consistent with the tidal model, or even
with simple geometric exclusion, and agrees with the result in \cite{jfmploff20} for nearest
neighbours.

\begin{figure}
\centerline{%
\includegraphics[width=0.80\textwidth,clip]{\figpath 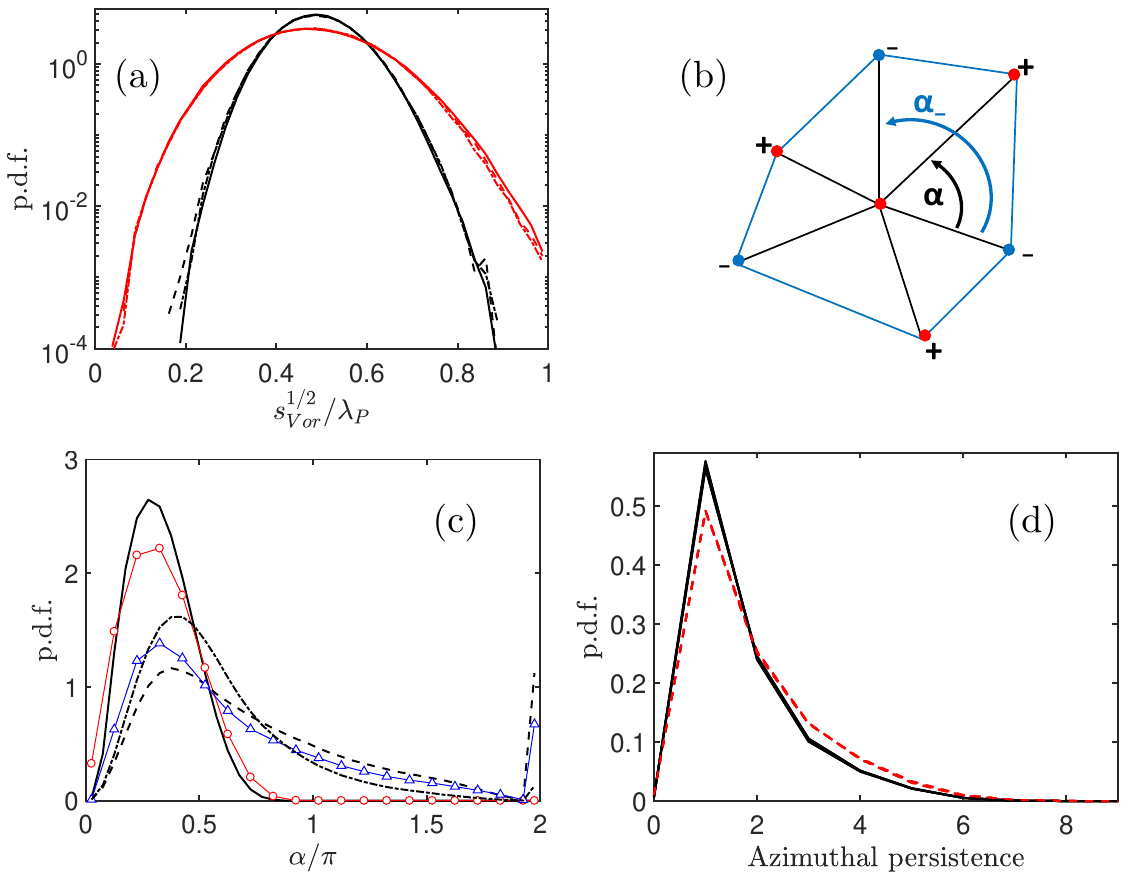}%
%
}%
\caption{%
(a) Probability density function of the diameter of the Voronoi tessellae. Black lines are
turbulence, and red lines are randomised vortex positions. Each p.d.f is the average of all
simulation times. \chndot, T768; \dashed, T1024; \solid, T1204a.
(b) Sketch of the definition of the angular distance between lattice neighbours, $\alpha$, and 
between neighbours of a given kind, $\alpha_\pm$.
(c) P.d.f. of the angular distance among the lattice neighbours to a given vortex. \solid, all
neighbours; \dashed, corrotating with the central vortex; \chndot, counterrotating;
\linecir, all neighbours, with randomised coordinates; \linetri, corrotating or
counterrotating neighbours, with randomised circulation. T1024a, averaged
over all times.
(d) P.d.f. of the length of the segments of mutually corrotating vortices in the crowns of
immediate neighbours to a given vortex. \solid, turbulence; \dashed, randomised circulation.
The figure includes all the cases in table \ref{tab:cases}, each one averaged over time, but
the results are indistinguishable at the plotting scale.
}
\label{fig:neighbour}
\end{figure}

Figure \ref{fig:neighbour} presents two more indicators of the tendency of turbulent flows to
be more organised than random point sets. Figure \ref{fig:neighbour}(a) present p.d.f.s of
the diameter, $s_{Vor}^{1/2}$, of the individual polygons in Voronoi tessellations, as in
figure \ref{fig:delaunay}(b). The distribution depends very little on the evolution time and
even on the flow parameters, and is shown in the figure averaged over the whole flow decay.
It follows from the area-covering property of tessellations, and from \r{eq:vortnum}, that
$\bra s_{Vor}\ket = \lambda_P^2/4$, or $\bra s_{Vor}^{1/2}\ket/ \lambda_P\approx
0.5$. This is true for the centres of gravity of turbulence and for random point sets, but
the standard deviation of the turbulence distributions $(0.08 \lambda_P)$ is much narrower
than for random sets $(0.125 \lambda_P)$.

The organisation of the vortices in the `crown' that forms the lattice neighbourhood of a
given vortex can be characterised by arranging them counterclockwise and measuring the angle
$\alpha$ among the consecutive Delaunay edges joining them with the central core (see sketch
in figure \ref{fig:neighbour}b). Since the average number of lattice neighbours is $\bra D
\ket=6$, one could expect $\bra\alpha\ket = 2\pi/\bra D\ket= \pi/3$. Empirically,
$\bra\alpha\ket/\pi=0.33\,(1\pm 0.5)$, which is essentially the same as for randomised vortex
positions. The situation is different for corrotating or counterrotation neighbours, for
which $\bra\alpha_\pm\ket/\pi \approx 2/\bra D_\pm\ket$ also holds, but the standard
deviation, $\bra\alpha^2_\pm\ket^{1/2}/ \bra\alpha_\pm\ket\approx 0.6$, is higher than for
the overall angles, and lower than for randomised vortex positions or circulations,
$\bra\alpha^2_\pm\ket^{1/2}/ \bra\alpha_\pm\ket\approx 0.7$. In fact, figure the
\ref{fig:neighbour}(c) shows that the distribution of the angular distance among
corrotating or counterrotating vortices is not just a shifted version of the p.d.f. for all the
vortices. Some of the Delaunay neighbours that corrotate (or counterrotate) with respect to
their reference vortex are immediate angular neighbours among themselves, and the modal value
of the angle distributions is close to the general mode. Their higher mean value is due to a
tail of larger angles, required for the total to add to $2\pi$. For example, the corrotating vortices
include a substantial spike at $\alpha_+=2\pi$, corresponding to crowns with a single
corrotating partner.

The sign of the circulation of the vortices within each Delaunay crown is slightly more
mixed than would correspond to a random distribution. Figure \ref{fig:neighbour}(d) shows
the p.d.f. of the length of the segments with similar sign around the crown. For example,
the sketch in figure \ref{fig:neighbour}(b) is maximally mixed, i.e., its highest azimuthal
persistence is one, because there are no contiguous vortices of the same sign. A minimally
mixed crown for a set of balanced vortices with the average $D= 6$ would have two segments
of length three. The dashed lines in figure \ref{fig:neighbour}(d) show that a set of random
points mostly has contiguous segments of length one or two, with a relatively fat tail
extending to full crowns of a single sign. The solid lines in the figure show that the
turbulence p.d.f. is slightly more peaked at unity than the random ones, with an average
persistence of 1.7. It describes a fairly inhomogeneous crown with less than two consecutive
vortices sharing the same sign, although with a slight tendency towards isolated corrotating
vortices. The latter is consistent with the already discussed tendency for corrotating
vortices to avoid each other.

\section{Crystals}\label{sec:crystals}

Before moving to our final attempt to extract ordered structures from the turbulent flow
fields, it is useful to explore whether what we have found up to now can be interpreted as
an approximation to stationary crystalline vortex lattices. Two-dimensional vortex crystals
are well known \cite{aref02}. Some of them are stable, and form spontaneously in
experiments. Forced two-dimensional turbulence is known to settle to stationary patterns
which are partly determined by the forcing and by the boundary conditions
\cite{fine95,jin00b,jim:gueg:07}, although many of the examples do not refer to the
Navier--Stokes equations, but to systems such as Bose-Einstein condensates, magnetised
plasmas, or superfluids, whose dissipation mechanisms are not exactly those of
Navier--Stokes fluids. Beautiful equilibrium vortex polygons have been observed in the polar
regions of planetary atmospheres \cite{taba:etal:20}.

\begin{figure}
\centerline{%
\includegraphics[height=0.23\textwidth,clip]{\figpath 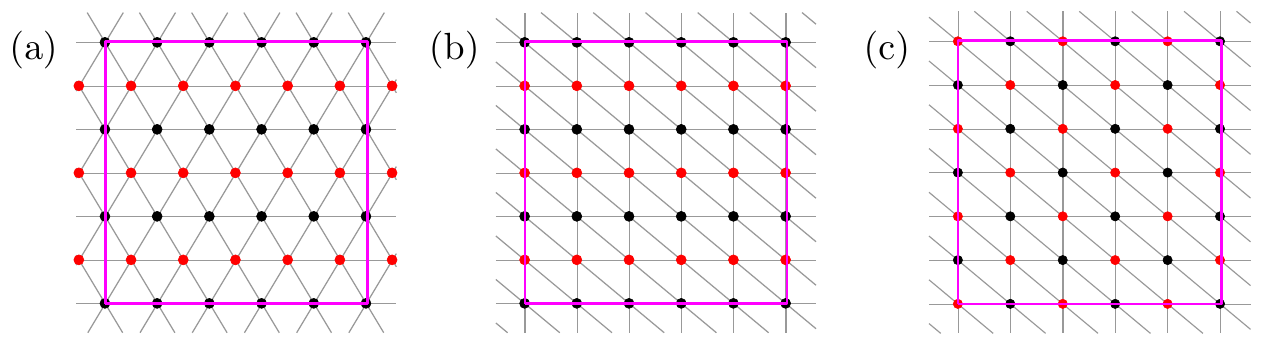}%
%
}%
\caption{%
(a) An approximately hexagonal lattice of staggered von K\'arm\'an vortex streets. Red dots
have positive circulation, and blue ones are negative.
(b) A rectangular lattice of aligned streets. 
(c) A rectangular chequerboard lattice. 
All these lattices are stationary equilibrium solutions.
}
\label{fig:karman}
\end{figure}

Most of these examples are regular arrangements of vortices of a single sign in a background
of opposite-sign vorticity, but mixed-sign vortex systems with zero overall circulation also
exist. The von K\'arm\'an vortex street is probably the best-known, and can be generalised
to a doubly periodic stationary lattice. Consider the two examples in figure \ref{fig:karman}(a,b).
There are two equilibrium configurations of a double street of point vortices,
both of which move with a constant velocity \cite[][\S 156]{lamb32}. In the first
one, two lines of vortices of opposite sign are staggered, and stacking an infinite number of these
`staggered' streets, as in figure \ref{fig:karman}(a), results in an approximately hexagonal
lattice. In the `aligned' equilibrium configuration of two vortex rows,
the vortices are vertically aligned with respect to each other, instead of staggered, and they
generalise to the `cubic' lattice in figure \ref{fig:karman}(b).

While the staggered and aligned von K\'arm\'an streets are the only known equilibrium
solutions for two lines of vortices of zero total circulation, there are many more doubly infinite
equilibrium lattices. Enumerating and analysing them is beyond the scope of this
paper, but an example is given in figure \ref{fig:karman}(c). Note that the three lattices
in figure \ref{fig:karman} are rotated versions of each other for particular aspect ratios
of order unity, but that they are generally distinct, and that they share a structure of
alternating layers of vorticity that induce high-velocity streams which could be used as a
model for the streams observed in  figure \ref{fig:flows}(c,d).

\begin{figure}
\centerline{%
\includegraphics[width=0.80\textwidth,clip]{\figpath 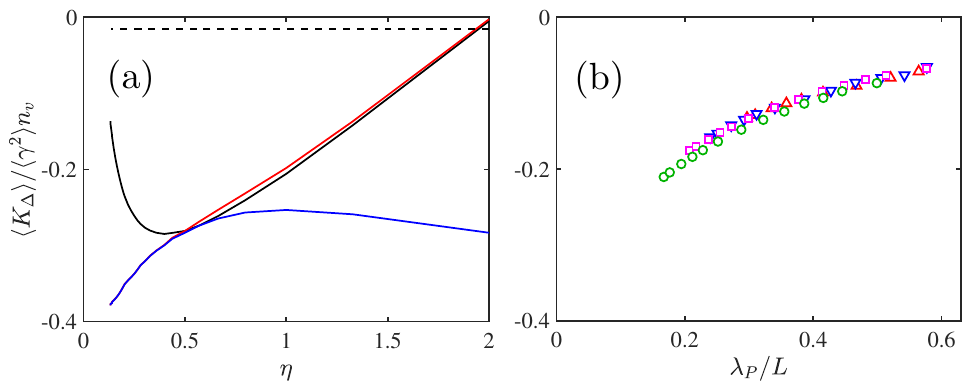}%
%
}%
\caption{%
(a) Normalised interaction energy for the three lattices in figure \ref{fig:karman}, versus
the aspect ratio. Black, staggered von K\'arm\'an lattice; red, aligned lattice; blue,
chequerboard. The dashed line is a random superposition of between $n_v=2$ and 100 Gaussian
cores in the computational box.
(b) Normalised interaction energy for the point-vortex approximation to the turbulence
simulations. Symbols as in table \ref{tab:cases}.
}%
\label{fig:lattener}
\end{figure}

The interaction energy of these lattices depends on the aspect ratio, defined as
$\eta=\Delta y/\Delta x$, where $\Delta x$ is the pitch within each row, and $\Delta y$ is
the distance between rows. This energy can be computed analytically in some cases
\cite{tkachen66a}, but it is most easily determined by approximating the lattice by a system
of compact cores, as in \S \ref{sec:energy}, and subtracting the self energy estimated in
appendix \ref{sec:lattice_ener}. This is done in figure \ref{fig:lattener}(a), where
energies are represented normalised per core, and shows that the regular lattices are
low-energy states with respect to randomised vortex distributions. The dashed line is the
normalised interaction energy for truly random vortex-core systems, computed in the same
way, and its small difference from zero is a measure of the accuracy of the estimates in
appendix \ref{sec:lattice_ener}.
 
Figure \ref{fig:lattener}(b) contains the energy for the point-vortex approximation to the
turbulence simulations in table \ref{tab:cases}, normalised as in figure
\ref{fig:lattener}(a), and shows that turbulence goes a good part of the way to minimising
its energy, reaching values comparable to those of infinite regular lattices. It would be
nice to relate this low energy to the stability of the different configurations, but
energy is not a good indicator of stability. The staggered von K\'arm\'an
double street is unstable for all aspect ratios except $\eta_c\approx 0.28055$, at which it is neutrally
stable to linear perturbations \cite{lamb32,jimenez87}, but its energy increases
monotonically with increasing $\eta$, without any special behaviour at $\eta_c$. The
lattices in figure \ref{fig:karman} have been drawn with an aspect ratio of order
$\eta=\tfrac{1}{2}$, suggested by isotropy. Figure \ref{fig:lattener}(a) shows that this is
close to the minimum energy configuration for the staggered von K\'arm\'an lattice in figure
\ref{fig:karman}(a), but this should not be taken as a stability criterion. It is known that
an equilateral triangular lattice is the only stable case for vortex lattices of a single
sign \cite{tkachen66}, but the stability of the mixed cases in figure \ref{fig:karman} is
unknown.

We are now in a position to discuss in which sense, if any, the statistics of turbulence 
approximate a lattice such as those in figure \ref{fig:karman}. Properties such as a
uniform Voronoi area (figure \ref{fig:neighbour}a) or a constant Delaunay degree (figure
\ref{fig:latt}a) are trivially satisfied by most regular lattices. The angle distribution in
figure \ref{fig:neighbour}(c) is not different enough from those of randomised vortices to
single out a particular lattice, but its narrow distribution around $\pi/3$ rules out the
symmetries in figures \ref{fig:karman}(b,c), which are closer to a bimodal distribution of
angles with peaks around $\pi/4$ and $\pi/2$. The azimuthal persistence in figure
\ref{fig:neighbour}(d), with approximately twice as many isolated vortices of one sign as
the number of corrotating pairs, is consistent with any of the layered lattices in figure
\ref{fig:karman}, but not with isolated vortices surrounded by crowns of opposite
circulation.

It is interesting that, although we have discussed the imbalance between corrotating and
counterrotating neighbours as indications of out-of-equilibrium behaviour, all the
equilibrium lattices in figure \ref{fig:karman} have twice as many counterrotating Delaunay
neighbours as corrotating ones. This is another consequence of the organisation into layers,
but is probably not enough to explain the results in figure \ref{fig:latt}. For one thing,
the observed imbalance in the number of neighbours of each class (3.5 to 2.5 in figure
\ref{fig:latt}b), is not close enough to 2 to unambiguously define a regular layered
lattice. The ratio between the average distance of the Delaunay neighbours to the distance
to the closest one (figure \ref{fig:latt}d), is more informative. This ratio is fairly close
to unity for any of the lattices in figure \ref{fig:karman}. Even for the square lattices in
figures \ref{fig:karman}(b,c), it is about $d_{av}/ d_{min}\approx 1.14$. The
average-to-minimum ratio for vortices of all classes in figure \ref{fig:latt}(d) is
approximately 1.8, much closer to the random value, 2.3, than to any regular grid. The set
of counterrotating vortices is intermediate between random and regular lattices, $d_{av}/
d_{min}\approx 1.54$, but the most `regular' distribution of distances is for the set of
vortices corrotating with the centre, for which $d_{av}/ d_{min}\approx 1.34$. Moreover,
figure \ref{fig:latt}(d) shows that the mean distance to the central vortex is very similar
for the three vortex classes, but that, in agreement with our previous analysis, it is the
tight pairs of corrotating vortices that are missing.

In summary, although the statistics presented up to now for the turbulence lattice are
consistent with local double layers of counterrotating vortices, they probably cannot be
used by themselves as indicators of an ordered lattice over long distances.
 
\section{Crystallography}\la{sec:stars}

We finally investigate whether the global vortex organisation can be characterised using
Fourier techniques inspired by X-ray crystallography. At first sight, this seems difficult.
Fourier analysis of turbulence is similar to powder crystallography, in which diffraction
takes place over disparate small crystals with random orientations, and whose typical
outcome is only a set of length scales \cite{sanders69}. There are two sources of randomness
in the Fourier analysis of turbulent flows. The first one is that individual fields have
random orientations, as in powders. The second is that the Fourier transform is an integral
over the whole computational box and, since we have seen that non-trivial vortex
arrangements probably only extend to local neighbourhoods, the transform also mixes
neighbourhoods with random orientations and positions. The randomisation due to the
orientation of individual fields is relatively easy to compensate, because we have access to
the spectrum of each field. The one due to the global transform requires substituting local
expansions for global ones.

In view of the difficulty of identifying global symmetries, we centre this section on
the simpler problem of determining whether some specific symmetry applies, on average, to
local neighbourhoods of individual vortices. We know from previous sections that the mean
distance between vortices is of the order of $\lambda_P$, and define neighbourhoods as disks
whose radius is some low multiple $R_P=\mu \lambda_P$. Consider the point-vortex
representation of the flow as a set of vortex centres located at $\bx_j$, with circulations
$\gamma_j$. To study the neighbourhood of the $i$-th vortex, define local polar coordinates,
$\bx_j-\bx_i \to ( r_{ij},\theta_{ij} )$, and construct
\beq
\widehat{\Gamma}^{[i]}_n = 
     \sum_j \gamma_{ij} \exp(-\ii n \theta_{ij}) ,
\la{eq:azim1}
\eeq
where $\gamma_{ij} = \gamma_i\gamma_j/|\gamma_i|$, and the sum extends over the vortex
centres satisfying $ r_{ij}\le R_P$. The effect of the sign factor $(\gamma/|\gamma|)$ in
$\gamma_{ij}$ is to classify vortices as co- or counter-rotating with the reference one,
$\bx_i$, rather than as positive or negative. Equation \r{eq:azim1} is essentially the
angular Fourier transform of the projection on the unit circle of the locations of the
vortices contained in a neighbourhood disk, and the complex $\widehat{\Gamma}^{[i]}_n$ can be
written as $|\widehat{\Gamma}^{[i]}_n| \exp(\ii\psi^{[i]}_n)$. To overcome the problem of
random orientations, we rotate the position of the vortices in each individual neighbourhood
so that some specified harmonic, $n_0$, is made real and positive,
\beq
(\widetilde{r}_{ij}, \widetilde{\theta}_{ij}) \to (r_{ij}, \theta_{ij}+\psi^{[i]}_{n_0}/n_0).
\la{eq:vorot}
\eeq
thus rotating other harmonics in \r{eq:azim1} to $\widehat{\Gamma}^{[i]}_n \exp(-\ii n
\psi^{[i]}_{n_0}/n_0)$, and $\widehat{\Gamma}^{[i]}_{n_0}$ to
$|\widehat{\Gamma}^{[i]}_{n_0}|$. Intuitively, this rotation aligns the average position of
the cloud of co- (counter-) rotating vortices to preferentially lie in one of the positive
(negative) lobes of $\cos(n_0\theta_{ij})$, and the resulting ensemble average is a
conditional average of the vortex lattice in these aligned coordinates.

\begin{figure}
\vspace*{2ex}%
\centerline{%
\includegraphics[width=0.95\textwidth,clip]{\figpath 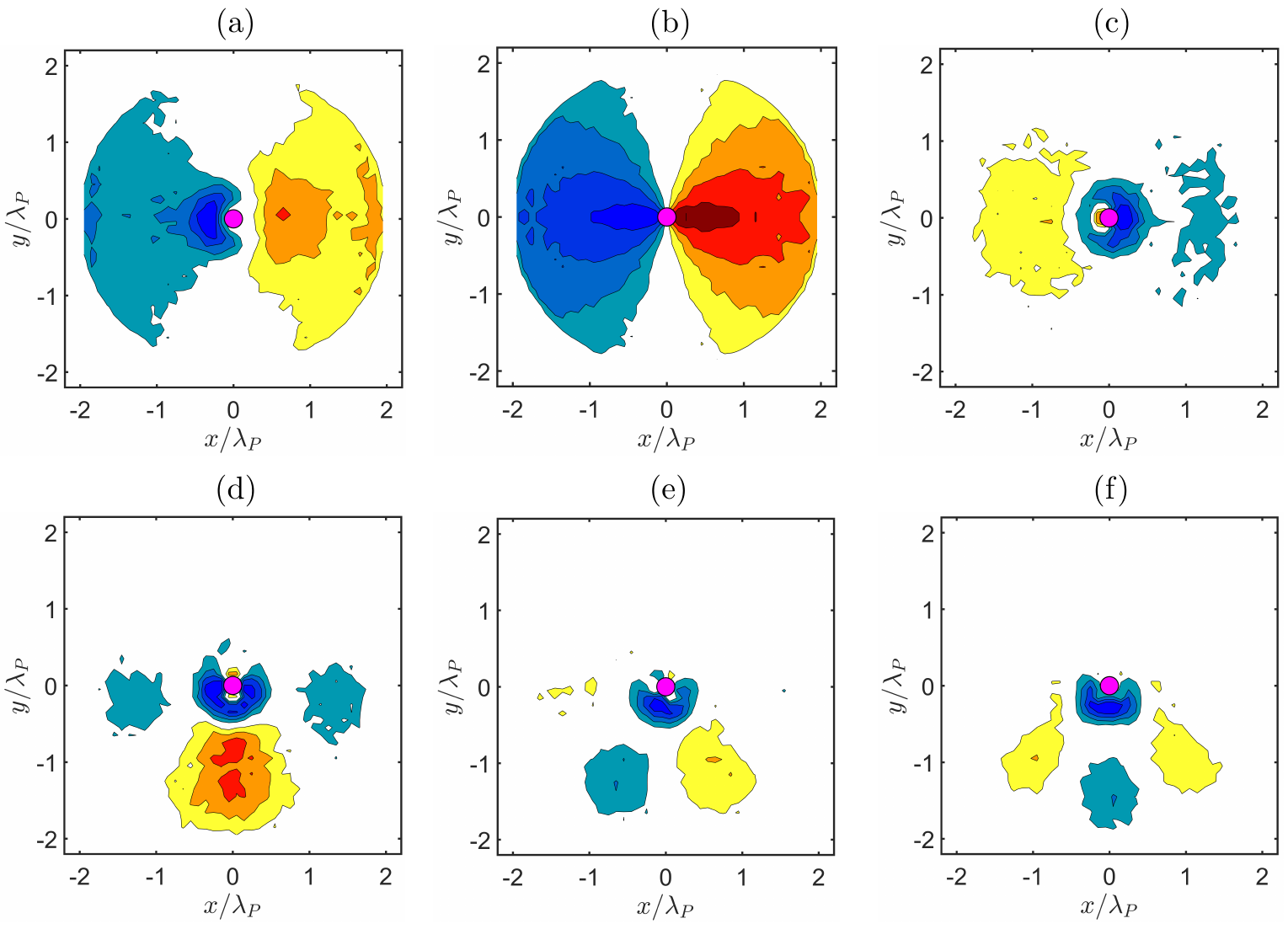}%
}%

\caption{%
Two-dimensional distribution of vortex positions in oriented flow neighbourhoods, normalised
to unit maximum absolute value. T1024a, as in figure \ref{fig:flows}. $\mu =2$. Contours
are $\pm[0.2:0.2:1]$. Red contours are for corrotating vortices, and blue are
counterrotating.
(a) Turbulence.
(b) Randomised positions.
(c-f) Turbulence minus randomised. (a-c) $n_0=1$. (d) $n_0=2$. (e) $n_0=3$. (f) $n_0=4$. All
figures lack the central reference vortex, located at $x=y=0$, that has been indicated by
the purple circle whose radius corresponds to the average vortex size in figure
\ref{fig:prims}(d).
}
\label{fig:vorpos}
\end{figure}

An example is figure \ref{fig:vorpos}(a), which presents results for an early time in the
evolution of T1024a, oriented to $n_0=1$, and compares it to a similar flow in which the
vortex positions have been randomised (figure \ref{fig:vorpos}b). These figures are signed
p.d.f.s in which each vortex is counted as $\pm 1$ according to whether it co- or
counter-rotates with the central vortex of the neighbourhood. Statistics are compiled over
2000 flow fields, which involve approximately $4\times 10^6$ vortices in $2\times 10^5$
neighbourhoods. The result is a relatively compact concentration of counterrotating vortices
to the left of the figure, $x\approx -0.4\lambda_P$, and a more diffuse distribution of
corrotating ones to the right. The randomised flow in figure \ref{fig:vorpos}(b) does not
show the asymmetry between co- and counter-rotating vortices. Its p.d.f simply reflects the
structure of the orienting cosine in \r{eq:azim1}. Figure \ref{fig:vorpos}(a) is essentially
a directionally unfolded version of the circulation correlation in figure
\ref{fig:energy1D}(c), but a better way to render it is probably figure \ref{fig:vorpos}(c),
where the turbulence and randomised distributions have been subtracted. This representation
is less affected by artefacts due to the orienting cosine, and becomes more informative when
applied to fields oriented to some of the higher harmonics, as in figures
\ref{fig:vorpos}(d-f). Missing from all the maps in figure \ref{fig:vorpos} is the central
reference vortex, corrotating with itself by default, which has been added as a disk at the
origin with the radius implied by the average vortex size in figure \ref{fig:prims}(d). The
compact counterrotating cloud present in all figures near the centre is the screening vortex
found in the correlations in \ref{fig:energy1D}(c). The compact dipole formed by these two
vortices appears to be the statistically preferred state of the vortices in turbulence, as
already seen in \S\ref{sec:energy}, \S\ref{sec:lattice} and \cite{jfmploff20}.

The more extended vortex clouds farther from the centre, with separations of the order of
$1.5\lambda_P$, are too far to be considered part of the local neighbourhood (see figure
\ref{fig:latt}c), and belong to the category of the `next closest' vortex. They are of the
size of the kinetic-energy streams in figure \ref{fig:flows}(c,d). An important parameter of
the analysis is the multiple $\mu \lambda_P$ that determines how many vortices are
considered to be part of the local neighbourhood disk. In general, larger neighbourhoods
result in a clearer asymmetry, although the signal deteriorates when $R_P\gtrsim L/2$ and
the conditioning interacts with the periodicity of the numerics, but the outer vortex clouds
change little as long as they fit within the analysis disk.

It is probably unwise to read too much in conditional averages, which always retain some
influence from the conditioning. The underlying symmetry in figure \ref{fig:vorpos} is
dictated by the harmonic used in \r{eq:vorot}. But the asymmetry between co- and
counter-rotating vortices in figures \ref{fig:vorpos}(a) and \ref{fig:vorpos}(c-f) is a flow
property, as shown by the comparison with figure \ref{fig:vorpos}(b). So is the separation
among the different vortex clouds in the figures, and it is tempting to see in figure
\ref{fig:vorpos} fragments of the regular grids in figure \ref{fig:karman}. Note that the
distances involved are rather large. We saw in \S\ref{sec:experiments} that a disk of radius
$\lambda_P$ approximately contains 15 vortices. The analysis disk used in figure
\ref{fig:vorpos}, with $\mu =2$, therefore represents the organisation of about 60
vortices.

\section{Conclusions}\label{sec:conclusions}

The results in this paper describe how decaying two-dimensional turbulence spontaneously
organises into a low-energy lattice of large vortices that is substantially more ordered
than a random vortex distribution, and evolves more slowly than other parts of the flow, but
which is neither fully ordered over long distances, nor stationary. It contains most of the
kinetic energy of the flow, and could perhaps be described as a vortex `liquid', rather than
a gas or a crystalline solid. We may call it a `stochastic crystal'.

We have shown that what could be called the `latent heat' of randomisation is not an issue
in turbulence decay. The vortices in the flow never explore the randomised states, and they
approximately conserve energy by staying within a very restricted fraction of the
configuration space of vortex positions. If we consider the process of vortex amalgamation
during decay as a late stage of the condensation of the flow vorticity into compact cores,
vortex formation and order generation are parts of the same process.

We have seen that maintaining a low relative energy of the vortex lattice requires mutual
screening of the far field of the vortex-induced velocity, but that this is not implemented
by attracting vortices of opposite sign to surround each other, as in electrostatics,
but by preferentially absorbing vortices of the same sign, in a process akin to tidal
disruption in strong gravitational fields.

The resulting vortex lattices are not organised enough to be considered equilibrium
solutions, but they are shown by various geometric analyses to locally approximate double
vortex layers, which move slowly and could perhaps be related to fixed points of the
dynamical system representing the flow. However, the equilibrium is constantly disrupted in
decaying turbulence by the capture process mentioned above, which ensures that the
neighbourhood of each vortex contains on average one more counterrotating vortex than
corrotating ones.

One of the referees called attention to the possible dependence of the above results
on the initial conditions. This is an important issue in transient or decaying flows. It is known,
for example, that the long-term behaviour of the large scales in decaying three-dimensional
turbulence depends on the small-wavenumber slope of the initial energy spectrum, $E_{qq}\sim
k^\zeta$, and that $\zeta=2$ and $\zeta=4$ are distinguished values that tend to persist during
the decay (see \cite{ishetal06} and references therein). This persistence is not preserved
in two dimensions, because the inverse cascade continuously pumps energy into the
large scales, where it `condenses' and destroys any pre-existing spectral structure
\cite{smithyak93}. All our simulations use $\zeta=-1/2$ \cite{jfmploff20}, but the shorter
spectral peak of case T1024a was specifically introduced to explore the effect of different
initial spectra, and a few cursory experiments were run with $\zeta=-2$ to $\zeta=4$ for
the same purpose. The main effect of $\zeta$ is to change the rate at which the large scales
condense, which also modifies the balance between disordered small vortices and ordered
large ones. Very `peaked' spectra $(\zeta>0)$ introduce an initial family of vortices of
approximately uniform size, which does not completely disappears by the time that the rest
of the energy has condensed. Tests like the distribution of Voronoi areas in figure
\ref{fig:neighbour}(a) suggest that the large vortices in these flows are even more
organised than in the cases discussed in the present paper. For an initially uniform enstrophy spectrum
$(\zeta=-2)$, the flow is condensed from the beginning, and vortex families do not have time
to form. The elucidation of the full range of possible behaviours would unfortunately extend
this paper beyond any reasonable length, and has not been attempted. The present results
should be seen as evidence of the spontaneous segregation of a relatively organised subset
of large vortices from an initial condition in which vortices have widely scattered (power law)
sizes and circulations.

Whether any of the mechanisms described here can be generalised to three-dimensional flows
is under investigation, but unknown at present.

\acknowledgments{%
This work was supported by the European Research Council under the Coturb grant
ERC-2014.AdG-669505.
}
  

%

\appendix
\section{Point-vortex simulations}\la{sec:pointv}

There are two compact vortex approximations used for comparison in the text. This appendix describes
them, and their numerical issues.
 
\subsection{Gaussian cores}\la{sec:gaussian}

The first approximation, used in \S\ref{sec:energy}--\ref{sec:crystals}, substitutes the precomputed turbulence
vortices by Gaussian cores, 
\beq
\omega_g(\bx)=(\gamma/\pi R^2) \exp(-|\bx-\bx_{cog}|^2/R^2),
\la{eq:gauss1}
\eeq
where $\gamma$ is the circulation of the turbulence vortex being approximated, and the
centre of gravity is defined over the vortex support, $s$, as
\beq
\bx_{cog}=s^{-1} \int_s \bx\,\omega(\bx) \dd^2\bx,
\la{eq:gauss1}
\eeq
The core radius is arbitrarily set to $R/L=10^{-2}$, except for some convergence tests at
$R/L=2.5\times 10^{-3}$, and the core area, when needed, is defined as $s_g=\pi R^2$.

There is no dynamics in this operation, which represents an existing flow field. The
cores are added to an empty $1024^2$ numerical grid, a uniform background vorticity is added
to zero the overall circulation, and flow quantities of this synthetic field are computed in the usual
manner.
 
The kinetic energy of the regular lattices in \S \ref{sec:crystals} are also computed using
this procedure.

\subsection{Hamiltonian point vortices}\la{sec:hamilton}

In contrast to the postprocessing operation just described, some simulations were run for
Hamiltonian systems of point vortices. Defining a complex variable, $z=x+\ii y$, each vortex
satisfies
\beq
\dr_t\olz_j=\olw_j=u_j-\ii v_j = \sum_{k\ne j} \frac{\gamma_k}{2\pi\ii (z_j-z_k)}, 
\la{eq:hamil1}
\eeq
where $\gamma_k$ is the circulation of the individual vortices, and the overline denotes complex
conjugation. For a doubly periodic flow in $L\times L$, the right-hand side is
substituted by  \cite{lamb32} 
\beq
\olw_j=\frac{1}{2\ii L }\sum_k \gamma_k \sum_{m'=\infty}^\infty \cot\left( \frac{\pi (z_j-z_k)}{L}+\ii m \pi\right), 
\la{eq:hamil2}
\eeq
where the inner sum includes $m=0$ except for $k=j$. In practice, this inner series can be
truncated to $m\le 2$. The system \r{eq:hamil2} is integrated with a second-order
Runge-Kutta, which is formally symplectic for uniform time step, $\Delta t$
\cite{sanzserna:94}. The spatial resolution is determined by this time step. The critical
case is when two corrotating vortices get closer than a distance $\ell$, in which case they
should rotate around each other with velocity $|w|=\gamma/2\pi\ell$. Accuracy requires that
$|w|\Delta t \ll \ell$, or $\ell\gg (\gamma\Delta t/2\pi)^{1/2}$. The accuracy constraint is
laxer for counterrotating pairs, which tend to translate in a straight line. For shorter
distances, corrotating and counterrotating pairs behave different for numerical reasons, and
there is an apparent correlation dip reminiscent of figure \ref{fig:energy1D}(c). This was
tested by changing the time step by a factor of 4, so that $\ell/\lambda_P=0.05-0.2$, and
became the reason why T1024a was rerun at a twice-shorter time step, without any effect.

\section{Vortex overlap}\la{sec:overlap}

Consider sets of $N$ circular vortices, of uniform diameter $R$, in a box of area $L^2$. Two
vortices intersect if their centres are at a distance $h<2R$, and the relevant parameter is
$\sigma =h/2R \in (0,1)$. The area of their intersection is
\beq
\Delta S/2 R^2=  \arccos(\sigma)-\sigma (1-\sigma^2)^{1/2}.
\la{eq:inter1}
\eeq
The probability density of the centre of a vortex being at a distance $h$ from another one
is $2\pi h/L^2$, so that the average intersection area is
\beq
\bra\Delta S_1\ket = \int_0^1 p(\sigma) \Delta S \dd\sigma = 
  \frac{16 \pi  R^4}{L^2} \int_0^1 \sigma \left [ \arccos(\sigma)-\sigma (1-\sigma^2)^{1/2}\right ]  \dd\sigma
  =  \frac{\pi^2  R^4}{L^2}  .
\la{eq:inter2}
\eeq
Since the total number of possible vortex pairs is $N(N-1)/2$, and the total vortex area in $\pi NR^2$, the 
relative area fraction of the intersections is
\beq
\bra\Delta S_N\ket/S_N =   (N-1)\pi  R^2/2L^2 \approx  0.5 S_N/L^2.
\la{eq:inter3}
\eeq
The experimental factor is closer to 0.45, but the relation holds very well, particularly
for the circular Gaussian cores.

\section{The vortex energy}\la{sec:lattice_ener}

Consider a system of $n_v$ compact vortices in a doubly periodic box of side $L$, assumed to
be of radius $R\ll L$ and divided into equal numbers of positive and negative circulation,
$\pm\gamma$. A rough estimation of the r.m.s. velocity fluctuations that they generate follows
from assuming their induced velocities to be independent variables whose individual
variance is the average
\beq
q_1'^2 = \frac{1}{L^2}\, \int_{L^2} \left( \frac{\gamma}{2\pi |\bx|} \right)^2  \dd^2 \bx \approx
     \frac{1}{L^2}\, \int_R^{\ell_0} \left( \frac{\gamma}{2\pi r} \right)^2  2\pi r\dd r =
     \frac{\gamma^2}{2\pi L^2}\,\log (\ell_0/R),
\la{eq:latt1}
\eeq
where $\ell_0$ is an outer scale beyond which the $1/r$ velocity law ceases to apply. The
expected velocity variance, or kinetic energy, due to $n_v$ cores is of the order of
\beq
q'^2 =q'^2_1 n_v \approx \frac{\gamma^2 n_v}{2\pi L^2} \log (\ell_0/R).
\la{eq:latt2}
\eeq
Note that the assumption in this equation, in the notation of the body of the paper, is that
ramdomised vortices only have self energy. If we similarly estimate the mean enstrophy as
\beq
\omega'^2 \approx (\gamma/R^2)^2 n_v (R/L)^2,
\la{eq:latt3}
\eeq
we obtain
\beq
q'^2/(\omega'^2 R^2) \sim \log (\ell_0/R).
\la{eq:latt4}
\eeq
For regular lattices such as those in \S \ref{sec:crystals},
$\ell_0=2(\Delta_1\Delta_2)^{1/2}/\pi$, where the $\Delta_j$ are the lengths of the two
vectors that define the lattice cell \cite{tkachen66a}. For a random superposition of
vortices in a periodic box, this becomes $\ell_0=2L/\pi$, and the logarithmic correction in
figure \ref{fig:energy1D}(b) applies. On the other hand, if the velocity is screened beyond
distances proportional to $R$, the correction disappears.

\end{document}